\documentclass[aps,preprint,onecolumn,superscriptaddress,showpacs,amssymb]{revtex4-1}
\usepackage[english]{babel}
\usepackage{graphics}
\usepackage{graphicx}
\usepackage{epsfig}
\usepackage{amssymb}
\usepackage{dcolumn}% Align table columns on decimal point
\usepackage{bm}% bold math
\usepackage{color}
\usepackage{natbib}
\usepackage{lineno}
\usepackage{sidecap}
\usepackage{verbatim}  % Needed for the "comment" environment to make LaTeX comments
\usepackage{vector}  % Allows "\bvec{}" and "\buvec{}" for "blackboard" style bold vectors in maths
\usepackage{wrapfig}
\usepackage{enumerate}
\usepackage{sidecap}
\usepackage{amsmath}
\usepackage{hyperref}

\begin{document}

\title{Anomalous band renormalization due to high energy $kink$ in the colossal thermoelectric material K$_{0.65}$RhO$_2$}

\author{Susmita Changdar}
\affiliation{Condensed Matter Physics and Material Sciences Department, S. N. Bose National Centre for Basic Sciences, Kolkata, West Bengal-700106, India}
\author{G. Shipunov}
\affiliation{Leibniz Institute for Solid State Research, IFW Dresden, D-01171 Dresden, Germany}
\author{N. B. Joseph}
\affiliation{Solid State and Structural Chemistry Unit, Indian Institute of Science, Bangalore, Karnataka-560012, India}
\author{N. C. Plumb}
\affiliation{Swiss Light Source, Paul Scherrer Institute,  CH-5232 Villigen PSI, Switzerland.}
\author{M. Shi}
\affiliation{Swiss Light Source, Paul Scherrer Institute, CH-5232 Villigen PSI, Switzerland.}
\author{B. B\"uchner}
\affiliation{Leibniz Institute for Solid State Research, IFW Dresden, D-01171 Dresden, Germany}
\author{Awadhesh Narayan}
\affiliation{Solid State and Structural Chemistry Unit, Indian Institute of Science, Bangalore, Karnataka-560012, India}
\author{S.\ Aswartham}
\affiliation{Leibniz Institute for Solid State Research, IFW Dresden, D-01171 Dresden, Germany}
\author{S.\ Thirupathaiah}
\email{setti@bose.res.in}
\affiliation{Condensed Matter Physics and Material Sciences Department, S. N. Bose National Centre for Basic Sciences, Kolkata, West Bengal-700106, India}
\date{\today}

\begin{abstract}
 We report on low-energy electronic structure and electronic correlations of K$_{0.65}$RhO$_2$,  studied using high-resolution angle-resolved photoemission spectroscopy (ARPES) technique and density functional theory (DFT) calculations. We observe a highly correlated hole pocket on the Fermi surface. We further notice that the correlations are momentum dependent. Most importantly, two $kinks$ at binding energies of 75 meV and 195 meV have been observed from the band dispersion in the vicinity of the Fermi level.  While the low energy $kink$ at 75 meV can be understood as a result of the electron-phonon interaction, the presence of high energy $kink$ at 195 meV is totally a new discovery of this system leading to an anomalous band renormalization. Based on systematic analysis of our experimental data,  we propose high frequency bosonic excitations as a plausible origin of the high energy anomaly. Further, we notice that the high energy anomaly has important implications in obtaining the colossal thermoelectric power of K$_{0.65}$RhO$_2$.

\end{abstract}

%\keywords{Suggested keywords}%Use showkeys class option if keyword
                              %display desired
\maketitle

%\tableofcontents

Strong electronic correlations are vital in yielding various exotic systems like, high-T$_c$ superconductors~\cite{Scalapino}, heavy fermionic materials\cite{Stewart1984}, quantum anomalous Hall insulators (QAHIs)~\cite{Liu2016}, half-metals~\cite{Groot1983}, Mott-insulators\cite{Mott1949}, itinerant magnets~\cite{Moriya1984}, and high thermopower materials\cite{MacDonald2006}. Electron-electron ($e$-$e$) correlations turn the materials to heavy fermionic systems, QAHIs, half-metals and Mott-insulators, while the electron-magnon interactions are expected to cause the itinerant ferromagnetism~\cite{Hertz1973, Edwards1973}. On the other, the electron-phonon ($e$-$ph$) interactions are thought to be playing a major role in the high-T$_c$ superconductivity~\cite{Lanzara2001} and high thermoelectricity~\cite{Wang2012}

For quite some time,  the compounds of the type A\textsubscript{x}BO\textsubscript{2} (A = Li, Na, and K, B = Co and Rh) have been the scientific topic of much interest due to their diverse physical properties~\cite{Elp1991, Terasaki1997, Sugiyama2006,  Shibasaki_2010}. Interestingly, depending on the amount of Na present in Na$_x$CoO$_2$, it exhibits superconductivity in the hydrated state for $x\approx$0.35~\cite{Schaak2003}, shows giant Seebeck coefficient for 0.7$<$x$<$1~\cite{Lee2006}, possesses magnetic ordering for $x\approx$0.75~\cite{Helme2005, Bayrakci2005} and charge ordering for $x\approx$0.5~\cite{Foo2004}. Crystal field splitting~\cite{Okazaki2011}, strong spin-orbit interactions~\cite{Koshibae2001}, electron-electron~\cite{Wang2003}, and  electron-phonon interactions~\cite{Donkov2008} are suggested for the cause of unusual physical properties in Na$_x$CoO$_2$. On the other hand, K$_x$RhO$_2$ a similar layered compound does not seem to be showing the physical properties as diverse as  Na$_x$CoO$_2$. K$_x$RhO$_2$ is reported to show a large thermoelectric power~\cite{Shibasaki_2010, Yao2012, Saeed2012}, but still is half to that of Na$_x$CoO$_2$~\cite{Terasaki1997},  despite both showing $e$-$ph$ interactions at a Debye frequency of 70-75 meV~\cite{Hasan2004, Yang2005, Chen2017}. Earlier ARPES report on K$_x$RhO$_2$ hinted at the importance of electron-boson scattering for the recorded high thermopowers in K$_x$RhO$_2$~\cite{Chen2017}. In this contribution, we study the effect of electronic correlation on the low-energy electronic structure of the layered  K$_{0.65(2)}$RhO$_2$ single crystal using ultra-high energy resolution ARPES technique and DFT calculations(see Experimental Methods section).
%to investigate the origin of relatively low thermopower in K$_x$RhO$_2$ compared to Na$_x$CoO$_2$.

\begin{figure}[ht]
  \centering
\includegraphics[width=\linewidth]{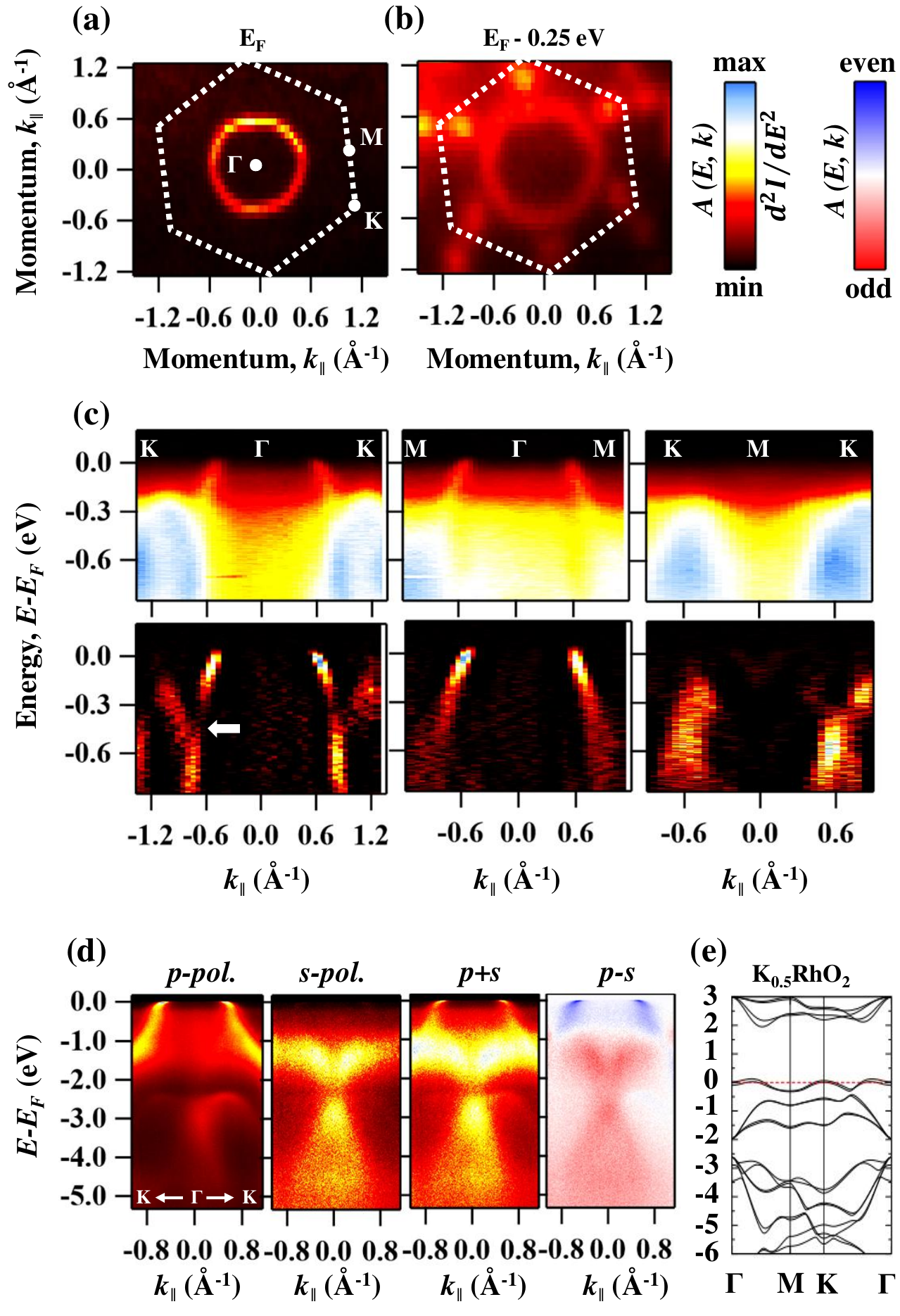}
\caption{ARPES measurements of K$_{0.65}$RhO$_2$. The data is measured using $p$-polarized light with a photon energy of 140 eV. (a) Fermi surface map. (b) Constant energy map taken at a binding energy of 0.25 eV below $E_F$. (c) Energy distribution maps (EDMs) showing the band dispersions along the $\Gamma$-$K$, $\Gamma$-$M$,  and $M-K$ high symmetry directions. Panels in (d) from left to right represent the Energy distribution maps measured with 60 eV photon energy using $p$-, $s$- polarized lights,  sum of $p$ and $s$, and difference between $p$ and $s$, respectively. (e) Energy-momentum plot of K$_{0.5}$RhO$_2$ obtained from the DFT calculations.}
\label{1}
\end{figure}

ARPES data of K$_{0.65}$RhO$_2$ are shown in Figure~\ref{1}. From the Fermi surface map shown in Fig.~\ref{1}(a), we observe one nearly circular-shaped Fermi pocket centred at $\Gamma$ with a Fermi vector of $k$=0.51$\pm$0.02~$\AA^{-1}$, covering almost 30\% of the total area of 2D hexagonal Brillouin zone (BZ). From the constant energy contour taken at a binding energy of 0.25 eV,  shown in Fig.~\ref{1}(b), we observe six tiny spectral sheets near six $K$ points. Moreover, at this binding energy, size of the Fermi pocket centred at $\Gamma$ has increased and the circular shape is turned into hexagonal shape. Energy distribution maps (EDMs) taken along the $\Gamma$-$M$, $\Gamma$-$K$, $K$-$M$ directions [see the top panels in Fig.~\ref{1}(c)] suggest that Fermi sheet centred at $\Gamma$ has holelike band dispersion. Further from the EDMs taken along the $\Gamma$-$K$ and $K$-$M$ directions, we realize that the tiny spectral sheet near the $K$ point is originated from another holelike band dispersion with a band-top at 0.25 eV below $E_F$. This is further confirmed from the 2$^{nd}$ derivative intensity (I) of the EDMs ($\frac{d^2I}{dE^2}$) as shown in the bottom panels of Fig.~\ref{1}(c).  Arrow on the 2$^{nd}$ derivative EDM in the $\Gamma$-$K$ direction  indicates an antiband crossing between the two holelike band dispersions at $\approx$ 0.4 eV below $E_F$. Above this binding energy, the two holelike bands are well separated in the momentum space. Importantly, no antiband crossing is found from the 2$^{nd}$ derivative EDM in the $\Gamma$-$M$ direction,  down to 0.8 eV of the binding energy.

As observed from the panels of Fig.~\ref{1}(d), EDMs taken along the $\Gamma$-$K$ direction,  the bands dispersing from the Fermi level down to 1 eV are only accessible with $p$-polarized light, while the bands below 1 eV are accessible with $s$-polarized light. We can conclude from this observation that the $a_{1g}$ band that is dominating in the vicinity of $E_F$ has even parity~\cite{Singh2000}, while the $e^{'}_{g}$ bands dominating below 1 eV  have the odd parity with the respect to our measuring geometry (see Experimental Methods section). By subtracting $s$-polarized data with the $p$-polarized as shown in $p-s$ panel of Fig.~\ref{1}(d),   we can clearly disentangle the even-parity states (blue colored)  from the odd parity states (red colored). Adding both the data of $p$- and $s$-polarized lights ($p+s$), we compare the experimental band dispersions with the DFT calculations along the $\Gamma$-$K$ direction shown in Fig.~\ref{1}(e). From this comparison we realize that  the experimental band dispersions qualitatively agree with the DFT calculations of K$_{0.5}$RhO$_2$ as shown in Fig.~\ref{1}(e). That means, the holelike band dispersions noticed from ARPES both at the $\Gamma$ and $K$ points are also reproduced from DFT.  However, while experimentally the top of holelike band near $K$ is at around 0.25 eV below $E_F$, the DFT calculations suggest that these bands cross $E_F$. Next,  comparing our experimental band structure with the available ARPES data on these type of systems,  analogous to Na$_x$CoO$_2$~\cite{Hasan2004, Yang2005, Arakane2011} and Li$_x$NaCo$_2$~\cite{Okamoto2017},  we could also observe only one circular-shaped hole pocket from the Fermi surface map. Most importantly, in agreement with the previous report on K$_{0.62}$RhO$_2$~\cite{Chen2017},  we identified an antiband crossing at $\approx$ 0.4 eV below $E_F$ in $\Gamma$-$K$ direction. Further, the top of holelike band at the $K$ point is found nearly at the same binding energy of 0.25 eV.

\begin{figure}[t]
\centering
\includegraphics[width=\linewidth]{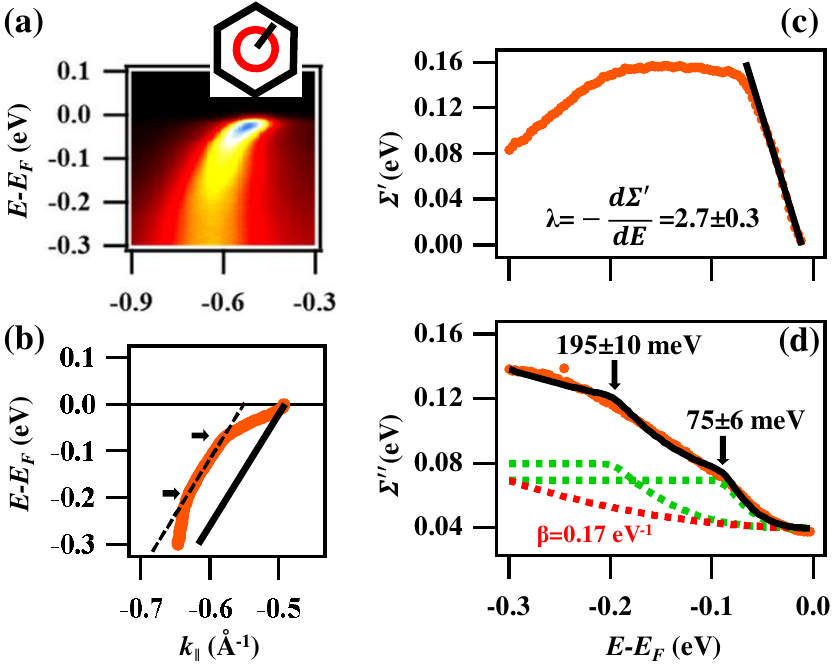}
\caption{(a) Energy distribution map taken along the $\Gamma$-$M$ orientation as shown in the inset. (b) Band dispersion extracted by fitting the momentum distribution curves of the  EDM shown in (a) using a Lorentzian function. Black dashed line in (b) is a linear fit to band dispersion at the higher binding energy within the window of (-0.2eV, -0.09eV). The arrows in (b) show the energy positions of the $kinks$.  (c) Imaginary part of the self-energy ($\Sigma^{''}$) extracted from the EDM shown in (a). In (c), the black curve represents fitting with combined functions of Fermi liquid theory-type and Eliasberg spectral functions (see the text). (d) Real part of the self-energy ($\Sigma^{'}$) extracted from the EDM shown in (a). In (a),  the black line is  linear fit to the data performed to extract the coupling constant $\lambda=2.7\pm0.3$.}
\label{2}
\end{figure}

Having thoroughly established the low energy electronic structure of K$_{0.65}$RhO$_2$, experimentally,  we then move on to the spectral function analysis of our experimental data. The band dispersion shown in Fig.~\ref{2}(b) is extracted from the EDM of Fig.~\ref{2}(a) by fitting the momentum distribution curves (MDCs) with a Lorentzian function,  analogous to the spectral function $A(E, k)=\frac{-1}{\pi}\frac{\Sigma^{''}}{(E_k-E_0-\Sigma^{'})^2+(\Sigma^{''})^2}$.
Here, $\Sigma^{'} (E)$ and $\Sigma^{''}(E)$ are the real and imaginary parts of the complex self-energy function defined as $\Sigma (E)=\Sigma^{'}+i\Sigma^{''}$. $E_k$ is the renormalized band dispersion which is generally obtained from the ARPES measurements [see Fig.~\ref{2}(b)] and $E_0$ is the bare band dispersion which is generally obtained by fitting the tight-binding parameters to the experimental data. Nevertheless, the bare band dispersion can also be obtained reasonably by fitting experimental data at higher binding energies where the electronic correlations are negligible. The black-dashed curve in Fig.~\ref{2}(b) is one of such fitting at the higher binding energies.  Then, the difference between $E_k$ [orange data in Fig.~\ref{2}(b)] and $E_0$ [solid black line in Fig.~\ref{2}(b), momentum offset to the dashed black line] provides the real part of the self-energy $\Sigma^{'} (E)=E_k-E_0$ as shown in Fig.~\ref{2}(d). On the other hand, the imaginary part of  self-energy shown in Fig.~\ref{2}(c) is calculated from the energy dependent spectral width [$\Delta k(E)$],  derived from the MDC fitting,  multiplied by the renormalized Fermi velocity ($v_F$=0.6 eV-\AA), $\Sigma^{''}=\Delta k(E) v_F$.

\begin{figure}[t]
\centering
\includegraphics[width=0.95\linewidth]{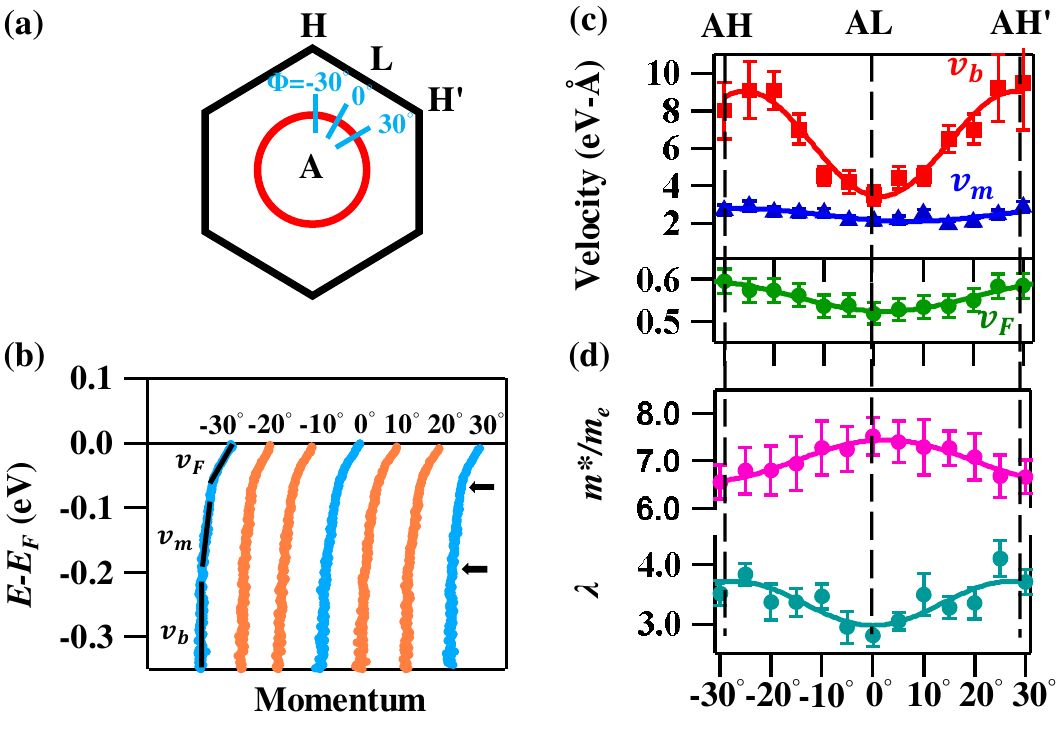}
\caption{(b) In-plane momentum dependant band dispersions extracted from the EDMs taken along the cuts by varying $\Phi$ as shown in (a). (c) In-plane momentum dependant Fermi and group velocities extracted by fitting with linear function within the binding energy windows as defined in (b). (d) In-plane momentum dependent coupling constant ($\lambda$) and carrier effective mass ($m^*/m_e$). In (c) and (d) the data are fitted with cosine functions.}
\label{3}
\end{figure}

Most interestingly,  we observe two $kinks$ from the band dispersion shown in Fig.~\ref{2}(b). These $kinks$ have direct implications on the imaginary and real parts of the self-energy as shown in Figs.~\ref{2}(c) and \ref{2}(d). Means, $\Sigma^{''}$ posses two humps corresponding to these two $kinks$. In order to understand origin of the hump, we performed a fitting to $\Sigma^{''}$ using multiple Eliashberg spectral functions [green dashed curves in Fig.~\ref{2}(c)] following the Debye model~\cite{Reinert}. The fitting resulted in two Debye frequencies 75$\pm$6 meV and 195$\pm$10 meV which are very much in agreement with the energy positions of the $kinks$ found from the band dispersion [Fig.~\ref{2}(b)]. In addition to multiple Eliashberg spectral functions, we needed to add a Fermi liquid-type spectral function [red dashed curve in Fig.~\ref{2}(c)], $\Sigma^{''} (E)=\alpha+\beta E^2$,  to properly fit $\Sigma^{''}$ for the binding energies beyond 0.2 eV. Here, $\alpha$ represents the spectral width due to impurity scattering and $\beta$ represents the strength of $e$-$e$ correlations. The derived $\beta$ value of 0.17$\pm0.03$ eV$^{-1}$ suggest week $e$-$e$ correlations in K$_{0.65}$RhO$_2$ compared to other high thermoelectric system having a $\beta$ value of 1.7 eV$^{-1}$~\cite{Nicolaou2010}.   We extracted a total coupling constant $\lambda$=2.7$\pm$0.3 by fitting $\Sigma^{'} (E)$ linearly with the formula of $\lambda=-\frac{d\Sigma^{'}}{dE}$ near $E_F$ as shown in Fig.~\ref{2}(c).  Interestingly, this value is quite high compared to the coupling constant ($\lambda$=0.4) reported earlier on  K$_{0.62}$RhO$_2$~\cite{Chen2017}. The differing coupling constants could have originated from the additional band renormalization due to the high energy anomaly at 195 meV. It is worth to mention here that such high value of coupling constants are also noticed from the high T$_c$ superconductors~\cite{Fink}. In Ref.~\cite{Fink},  a total coupling constant of $\lambda=3.9$ has been reported for (Bi,Pb)$_2$Sr$_2$CaCu$_2$O$_{8+\delta}$ (Bi2212), with the $e$-$ph$ interaction contribution of 2.3 and band renormalization contribution of 1.6. Following the same analogy, we subtracted the band renormalization contribution ($\lambda_b=\frac{v_b}{v_m}-1$=1.2$\pm$0.1) from the total coupling constant to obtain an $e$-$ph$ coupling constant of $\lambda_{e-ph}$=1.5$\pm$0.4. This value, within the error-bars, is in good agreement with the coupling constant independently obtained from the imaginary part of self-energy~\cite{Fink}, $\lambda_{e-ph}=\frac{2 \Sigma^{''}(-\infty)}{\Omega_0 \pi}$=1.19$\pm$0.1. Here $\Sigma^{''}(-\infty)$=140 meV and $\Omega_0$=75$\pm$6 meV.   Further in supporting our observation, a coupling constant of $\lambda_{e-ph}$=1 has been estimated for the multiboson-electron scattering in case of the misfit cobaltate, [Bi$_2$Ba$_2$O$_4$][CoO$_2$]$_2$~\cite{Nicolaou2010} which has the identical CoO$_2$ slabs to those in Na$_{x}$CoO$_2$.

 %choice of higher binding energy window considered for finding bare band dispersion ($E_0(k)$). In order to simulate the bare band dispersion, in this study we considered the energy range between -0.3 and -0.2 eV, whereas in Ref.~\onlinecite{Chen2017} the energy range is taken between -0.2 and -0.18 eV. Therefore in Ref.~\onlinecite{Chen2017}, the high energy (HE) $kink$ at 195 meV is not taken into account while estimating the coupling constant.

\begin{figure*}[htbp]
\centering
\includegraphics[width=\linewidth]{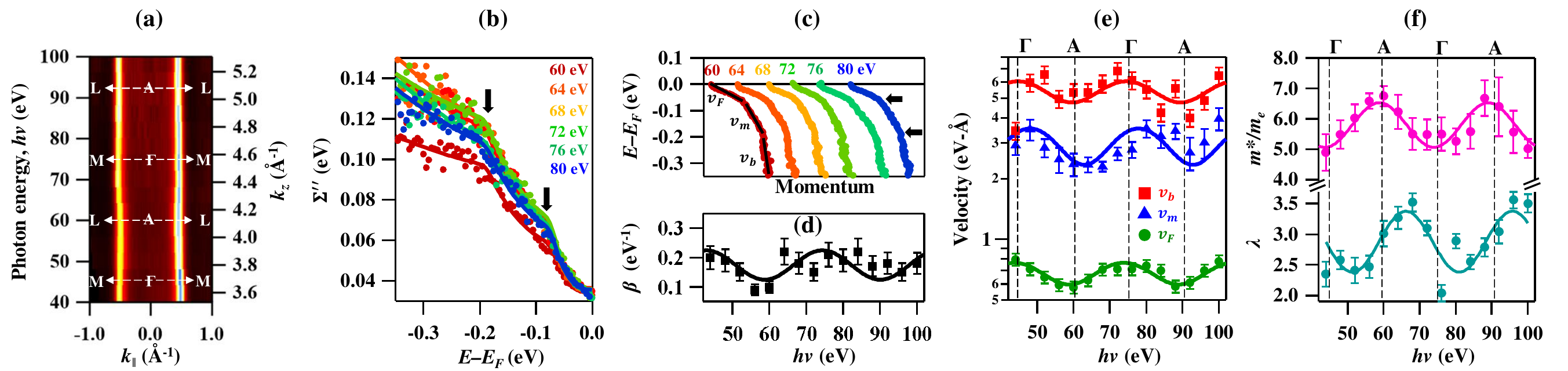}
\caption{(a) Out-of-plane Fermi surface map taken in the $k_z-k_{\|}$ plane. (b) Representative $k_z$ ($h\nu$) dependent imaginary part of the self-energy $\Sigma^{''}$. (c) Representative $k_z$ dependent dependant band dispersions. (d) $k_z$ dependent strength of electron-electron correlation extracted by fitting with Fermi liquid-type spectral function (see the text). (e) $k_z$ dependent Fermi and group velocities extracted by fitting with linear function within the binding energy window as shown in (c). (f) $k_z$ dependent coupling constant and carrier effective mass. In (d) and (e) the data are fitted with cosine functions.}
\label{4}
\end{figure*}

In-plane electronic correlations are evaluated for the $AHL$ plane as shown in Figure~\ref{3}.  Fig.~\ref{3}(b) depicts representative  band dispersions taken along the cuts $AH$, $AL$, and $AH'$ by rotating $\Phi$=-30$^\circ$ to 30$^\circ$ as demonstrated in Fig.~\ref{3}(a). From Fig.~\ref{3}(b),  we can again observe multiple $kinks$ from the all band dispersions almost at the same binding energy of 75 meV and 195 meV. We also estimated Fermi velocity $v_F$, and group velocities, $v_m$ and $v_b$, by fitting the band dispersions linearly in the binding energy windows (0, 0.07) eV, (0.1, 0.2) eV, and (0.21, 0.3) eV, respectively. The calculated Fermi and group velocities vary sinusoidally with $\Phi$ as shown in Fig.~\ref{3}(c),  having maximum Fermi and group velocities, $v_F$=0.6$\pm$0.03 eV-$\AA$, $v_m$=2.9$\pm$0.5 eV-$\AA$, $v_b$=9.1$\pm$1 eV-$\AA$ along the $AH$ direction and minimum Fermi and group velocities, $v_F$=0.52$\pm$0.02 eV-$\AA$, $v_m$=2.1$\pm$0.3 eV-$\AA$, $v_b$=3.5$\pm$0.5 eV-$\AA$ along the $AL$ direction. These observations are consistent with the reported $\Phi$ dependent Fermi velocities of Na$_x$CoO$_2$~\cite{Geck2007}. The total coupling constant ($\lambda$) and the effective mass of the hole pocket,  $m^*$ = $\frac{\hbar k_F}{v_F}$,  as a function of $\Phi$ are plotted in Fig.~\ref{3}(d). From Fig.~\ref{3}(d),  we can notice that the coupling constant is maximum ($\lambda$=3.7) for the electrons dispersing along $AL$ and is minimum ($\lambda$=2.9) for the electrons dispersing along $AH$. Further, the effective mass is minimum ($m^*$ = 6.5 $m_e$) along $AH$ and is maximum ($m^*$ = 7.4 $m_e$) along $AL$.

Next, electronic correlations have been evaluated for the out-of-plane momentum ($k_z$) direction as shown in Figure~\ref{4}. Fig.~\ref{4}(a) depicts the Fermi surface (FS) map taken in the $k_z-k_{\|}$ plane by varying the photon energy ($h\nu$) between 40 and 100 eV insteps of 4 eV. As can be seen from the $k_z$ FS map,  no change in the Fermi vector is noticed along the $k_z$ direction, suggesting a nearly 2D hole pocket without electron hopping in the $k_z$ direction. In Fig.~\ref{4}(b) we show representative imaginary part of the self-energy extracted from the EDMs measured with varying photon energies ($k_z$ dependent). From each photon energy data, we consistently observe two humps in  $\Sigma^{''} (E)$, within the error-bars almost at the same binding energies of 75 meV and 195 meV. This is in very good agreement with $kinks$ observed from the band dispersions extracted from corresponding photon energies [see Fig.~\ref{4}(c)]. As discussed earlier, we could reasonably fit $\Sigma^{''}$ at every photon energy using combined double-Eliashberg and Fermi liquid-type spectral functions as shown in Fig.~\ref{4}(b). We estimated $\beta$ from the fittings and is plotted as a function of $k_z$ (h$\nu$) as shown in Fig.~\ref{4}(d). From Fig.~\ref{4}(d),  we can notice that the $e-e$ correlations hardly change along $k_z$ (within the error-bars).  The estimated Fermi and group velocities are plotted in Fig.~\ref{4}(e) as a function of $k_z$. From Fig.~\ref{4}(e), we can notice that all the velocities vary sinusoidally with $k_z$ having minima at 60 and 92 eV and maxima at 43 and 75 eV photon energies. By considering the inner potential V$_0$=12$\pm$2 eV and using the formula $k_z~=~\sqrt{\frac{2m}{\hbar^2} (V_0 + E_k)}$,  we identify that the photon energies 60 and 92 eV extract the bands from the $AHL$ plane and the photon energies 45 and 75 eV extract the bands from the $\Gamma MK$ plane. Thus from Fig.~\ref{4}(e), we can find that the Fermi velocity is minimum at the $A$ point ($v_F$=0.6$\pm$0.04 eV-$\AA$) and is maximum at the $\Gamma$ point ($v_F$=0.76$\pm$0.06 eV-$\AA$).  Similarly, the group velocities $v_m$ and $v_b$ are minimum at $A$ (2.3$\pm$0.3, 4.76$\pm$0.5) eV-$\AA$ and are maximum at $\Gamma$ (3.53$\pm$0.5, 6.06$\pm$0.8) eV-$\AA$. With the help of Fermi velocity and Fermi momentum,  we estimated the effective mass of the hole pocket and plotted them as a function of $k_z$ as shown in Fig.~\ref{4}(f). A maximum effective mass is realized ($m^*$ = 6.51 $m_e$) at $A$, while a minimum effective mass is realized ($m^*$ = 5.06 $m_e$) at $\Gamma$. Further, the $k_z$ dependent total coupling constants are plotted in Fig.~\ref{4}(f). Note here that the maximum ($\lambda$=3.37) and minimum ($\lambda$=2.38) coupling constants are shifted by h$\nu$=5 eV from the photon energy positions of the high symmetry points, while still the photon energy difference between the two extrema is invariant ($\approx$ 15 eV).

Since we completely extracted the in-plane and the out-of-plane Fermi sheets using ARPES, with the help of Luttinger's theorem~\cite{Luttinger1960}, we are able to estimate the hole carrier density $n_h$=0.3$\pm$0.03 per unit cell. This value is in very good agreement with the $K$  deficiency percentage of the measured sample K$_{0.65(2)}$RhO$_2$ (1-$x$=0.35$\pm$0.02) from the stoichiometric KRhO$_2$. Thus, the ARPES data confirm EDAX estimate of the chemical composition. As clearly demonstrated from our ARPES data,  K$_{0.65}$RhO$_2$ possess two $kinks$. While the $kink$ at 75 meV is consistent with the previous studies of Raman spectroscopy showing active $E_{1g}+E_{2g}+A_{1g}$ Raman modes at around 500 cm$^{-1}$ from K$_{0.63}$RhO$_2$ ~\cite{Zhang2015}, the high energy $kink$ at 195 meV is totally new a finding of this study. Though  the origin of low energy $kink$ is reasonably understood, the origin of HE $kink$ is yet to be established.   So far existing ARPES studies on these systems did not concentrate on the electronic correlations beyond 0.2 eV binding energy. Therefore, we are unable to compare the HE $kink$ directly with previous ARPES studies of these systems. Nevertheless,  as can be seen from Figs.~\ref{2} and ~\ref{4}, we can reasonably fit $\Sigma^{''} (E)$ with multiple Debye frequencies at 75 meV and 195 meV. This suggests a plausible phononic origin for the HE $kink$. In fact, such an observation of bosonic scattering at higher energies has been noticed from Fe (100) at $\approx$ 160 meV~\cite{Schaefer2004}, graphene at $\approx$ 200 meV~\cite{Mazzola2013}, and cuprates at $\approx$ 350-400 meV below $E_F$~\cite{Valla2007, Inosov2007}.  The other existing mechanisms for the HE anomaly are the matrix element effects~\cite{Basak2009, Rienks2014} and spin-fluctuations~\cite{Graf2007}. As observed in this study and reported in the literature, near the Fermi level only one band disperses from $E_F$ down to a binding energy of 0.4 eV~\cite{Ou2009, Chen2017}. The same has been confirmed from the DFT calculations as well, especially, in $AHL$ plane~\cite{Singh2000}.  Since the observed HE $kink$ is at around 195 meV and only one band dispersion present within this energy range, it is highly unlikely that the HE $kink$ originated from the matrix elements. Further, the spin-fluctuations origin can be negated as the transport properties of K$_{0.65}$RhO$_2$ are nearly insensitive to the applied magnetic fields down to the lowest possible temperature~\cite{Zhang2014}. Finally, as demonstrated in  Fig.~\ref{1} the antiband crossing occur at $\approx$ 0.4 eV below $E_F$ which shows no effect on the $kink$ at 195 meV, ruling out the band structure origin as well.  Hence, the only convincing mechanism for the HE $kink$ must be the electron-boson scattering at higher frequencies. But the present available literature on these systems is insufficient to confirm the same.

Our estimate of average Fermi velocity over the entire BZ $v_F$=0.62$\pm$0.04 eV-$\AA$ is far less than the Fermi velocity ($v_F$=0.96$\pm$0.02 eV-$\AA$) reported earlier on K$_{0.62}$RhO$_2$~\cite{Chen2017}. On the other hand,  the average carrier effective mass estimated from this study, $m^*$=6.44$m_e$ is a factor of 4.7 less than the effective mass reported for Na$_{x}$CoO$_2$~\cite{Hasan2004}. From this,  we can conclude that K$_{0.65}$RhO$_2$ is relatively less correlated compared to Na$_{x}$RhO$_2$, but more correlated than what was thought earlier~\cite{Chen2017}. With the help of average Fermi vector ($k_F$=0.51$\pm$0.02 $\AA^{-1}$) and Fermi velocity,  we estimated the Seebeck coefficient using the Boltzman theory~\cite{Singh2000, MacDonald2006},  $ S=\frac{2\pi^2 k^{2}_{B} T}{3 e k_F v_F}$, of 46$\pm$5 $\mu$V/K at T=300 K. This value is in excellent agrement with the Seebeck coefficient $S_{300 K}$=46.3 $\mu$V/K derived from the transport measurements on K$_{0.63}$RhO$_2$~\cite{Yao2012}. We further verified the validity of Boltzman theory in the present context by evaluating the Seebeck coefficient for Na$_{0.65}$CoO$_2$. Considering $k_F$=0.6 $\AA^{-1}$ and averaged $v_F$=2.75 eV-$\AA$ from Ref.~\cite{Arakane2011}, we estimated the coefficient $S_{300K}$=89 $\mu$V/K which is in very good agrement with the value of $\approx$ 90 $\mu$V/K obtained from the transport measurements on Na$_{0.67}$CoO$_2$~\cite{Pandiyan2013} and with the value of $\approx$ 85 $\mu$V/K obtained from DFT calculations on Na$_{0.67}$CoO$_2$~\cite{Singh2007}. Thus, the Boltzman theory is sufficient to understand the enhanced thermoelectric power in these systems.

In conclusion, we systematically studied the low-energy electronic structure of  K$_{0.65}$RhO$_2$ using ARPES technique and DFT calculations.  We observe a correlated hole pocket centred at the $\Gamma$ point with a Fermi velocity of $v_F$=0.76 eV-$\AA$.  In going from $\Gamma$ to $A$ the strength of $e$-$ph$ scattering, represented by the coupling constant ($\lambda$), increases by a factor of 1.41 and varies sinusoidally along the $k_z$ momenta. We further notice relatively stronger correlations along the $AH$ direction ($\lambda=3.7$) compared to the $AL$ direction ($\lambda=2.9$). These in-plane and out-of-plane momenta dependent coupling constants follow the nature of momentum dependent Fermi velocities. Further, we observe multiple $kinks$ from the band dispersion at 75 meV and 195 meV. While the low energy $kink$ (75 meV) is well understood as the result of $e$-$ph$ scattering, the observation of high energy anomaly at 195 meV is a new discovery of this study and we propose high frequency boson scattering as the origin.

S.C. acknowledges University Grants Commission (UGC), India for the PhD fellowship. N.B.J. acknowledges support from the Prime Minister's Research Fellowship. A.N. acknowledges support from the startup grant (SG/MHRD-19-0001) at the Indian Institute of Science. S.A. and B.B. acknowledges financial support from Deutsche Forschungsgemeinschaft (DFG), Germany through the project nos.  419457929 and 405940956. S.T. acknowledges financial support by Department of Science and Technology (DST), India through the INSPIRE-Faculty program (Grant No. IFA14 PH-86). S.T. acknowledges the financial support given by SNBNCBS through the Faculty Seed Grants program. Authors thank the Department of Science and Technology, India (SR/NM/Z-07/2015) for the financial support and Jawaharlal Nehru Centre for Advanced Scientific Research (JNCASR) for managing the project. Authors thank E.D.L. Reinks for fruitful discussions.

\bibliography{KRhO2}

%merlin.mbs apsrev4-1.bst 2010-07-25 4.21a (PWD, AO, DPC) hacked
%Control: key (0)
%Control: author (8) initials jnrlst
%Control: editor formatted (1) identically to author
%Control: production of article title (-1) disabled
%Control: page (0) single
%Control: year (1) truncated
%Control: production of eprint (0) enabled
\begin{thebibliography}{51}%
\makeatletter
\providecommand \@ifxundefined [1]{%
 \@ifx{#1\undefined}
}%
\providecommand \@ifnum [1]{%
 \ifnum #1\expandafter \@firstoftwo
 \else \expandafter \@secondoftwo
 \fi
}%
\providecommand \@ifx [1]{%
 \ifx #1\expandafter \@firstoftwo
 \else \expandafter \@secondoftwo
 \fi
}%
\providecommand \natexlab [1]{#1}%
\providecommand \enquote  [1]{``#1''}%
\providecommand \bibnamefont  [1]{#1}%
\providecommand \bibfnamefont [1]{#1}%
\providecommand \citenamefont [1]{#1}%
\providecommand \href@noop [0]{\@secondoftwo}%
\providecommand \href [0]{\begingroup \@sanitize@url \@href}%
\providecommand \@href[1]{\@@startlink{#1}\@@href}%
\providecommand \@@href[1]{\endgroup#1\@@endlink}%
\providecommand \@sanitize@url [0]{\catcode `\\12\catcode `\$12\catcode
  `\&12\catcode `\#12\catcode `\^12\catcode `\_12\catcode `\%12\relax}%
\providecommand \@@startlink[1]{}%
\providecommand \@@endlink[0]{}%
\providecommand \url  [0]{\begingroup\@sanitize@url \@url }%
\providecommand \@url [1]{\endgroup\@href {#1}{\urlprefix }}%
\providecommand \urlprefix  [0]{URL }%
\providecommand \Eprint [0]{\href }%
\providecommand \doibase [0]{http://dx.doi.org/}%
\providecommand \selectlanguage [0]{\@gobble}%
\providecommand \bibinfo  [0]{\@secondoftwo}%
\providecommand \bibfield  [0]{\@secondoftwo}%
\providecommand \translation [1]{[#1]}%
\providecommand \BibitemOpen [0]{}%
\providecommand \bibitemStop [0]{}%
\providecommand \bibitemNoStop [0]{.\EOS\space}%
\providecommand \EOS [0]{\spacefactor3000\relax}%
\providecommand \BibitemShut  [1]{\csname bibitem#1\endcsname}%
\let\auto@bib@innerbib\@empty
%</preamble>
\bibitem [{\citenamefont {Scalapino}()}]{Scalapino}%
  \BibitemOpen
  \bibfield  {author} {\bibinfo {author} {\bibfnamefont {D.~J.}\ \bibnamefont
  {Scalapino}},\ }in\ \href {\doibase 10.1201/9780203737965-10} {\emph
  {\bibinfo {booktitle} {Superconductivity}}}\ (\bibinfo  {publisher}
  {Routledge})\ pp.\ \bibinfo {pages} {449--560}\BibitemShut {NoStop}%
\bibitem [{\citenamefont {Stewart}(1984)}]{Stewart1984}%
  \BibitemOpen
  \bibfield  {author} {\bibinfo {author} {\bibfnamefont {G.~R.}\ \bibnamefont
  {Stewart}},\ }\href {\doibase 10.1103/RevModPhys.56.755} {\bibfield
  {journal} {\bibinfo  {journal} {Rev. Mod. Phys.}\ }\textbf {\bibinfo {volume}
  {56}},\ \bibinfo {pages} {755} (\bibinfo {year} {1984})}\BibitemShut
  {NoStop}%
\bibitem [{\citenamefont {Liu}\ \emph {et~al.}(2016)\citenamefont {Liu},
  \citenamefont {Zhang},\ and\ \citenamefont {Qi}}]{Liu2016}%
  \BibitemOpen
  \bibfield  {author} {\bibinfo {author} {\bibfnamefont {C.-X.}\ \bibnamefont
  {Liu}}, \bibinfo {author} {\bibfnamefont {S.-C.}\ \bibnamefont {Zhang}}, \
  and\ \bibinfo {author} {\bibfnamefont {X.-L.}\ \bibnamefont {Qi}},\ }\href
  {\doibase 10.1146/annurev-conmatphys-031115-011417} {\bibfield  {journal}
  {\bibinfo  {journal} {Annual Review of Condensed Matter Physics}\ }\textbf
  {\bibinfo {volume} {7}},\ \bibinfo {pages} {301} (\bibinfo {year}
  {2016})}\BibitemShut {NoStop}%
\bibitem [{\citenamefont {de~Groot}\ \emph {et~al.}(1983)\citenamefont
  {de~Groot}, \citenamefont {Mueller}, \citenamefont {Engen},\ and\
  \citenamefont {Buschow}}]{Groot1983}%
  \BibitemOpen
  \bibfield  {author} {\bibinfo {author} {\bibfnamefont {R.~A.}\ \bibnamefont
  {de~Groot}}, \bibinfo {author} {\bibfnamefont {F.~M.}\ \bibnamefont
  {Mueller}}, \bibinfo {author} {\bibfnamefont {P.~G.~v.}\ \bibnamefont
  {Engen}}, \ and\ \bibinfo {author} {\bibfnamefont {K.~H.~J.}\ \bibnamefont
  {Buschow}},\ }\href {\doibase 10.1103/PhysRevLett.50.2024} {\bibfield
  {journal} {\bibinfo  {journal} {Phys. Rev. Lett.}\ }\textbf {\bibinfo
  {volume} {50}},\ \bibinfo {pages} {2024} (\bibinfo {year}
  {1983})}\BibitemShut {NoStop}%
\bibitem [{\citenamefont {Mott}(1949)}]{Mott1949}%
  \BibitemOpen
  \bibfield  {author} {\bibinfo {author} {\bibfnamefont {N.~F.}\ \bibnamefont
  {Mott}},\ }\href {\doibase 10.1088/0370-1298/62/7/303} {\bibfield  {journal}
  {\bibinfo  {journal} {Proceedings of the Physical Society. Section A}\
  }\textbf {\bibinfo {volume} {62}},\ \bibinfo {pages} {416} (\bibinfo {year}
  {1949})}\BibitemShut {NoStop}%
\bibitem [{\citenamefont {Moriya}\ and\ \citenamefont
  {Takahashi}(1984)}]{Moriya1984}%
  \BibitemOpen
  \bibfield  {author} {\bibinfo {author} {\bibfnamefont {T.}~\bibnamefont
  {Moriya}}\ and\ \bibinfo {author} {\bibfnamefont {Y.}~\bibnamefont
  {Takahashi}},\ }\href {\doibase 10.1146/annurev.ms.14.080184.000245}
  {\bibfield  {journal} {\bibinfo  {journal} {Annual Review of Materials
  Science}\ }\textbf {\bibinfo {volume} {14}},\ \bibinfo {pages} {1} (\bibinfo
  {year} {1984})}\BibitemShut {NoStop}%
\bibitem [{\citenamefont {MacDonald}(2006)}]{MacDonald2006}%
  \BibitemOpen
  \bibfield  {author} {\bibinfo {author} {\bibfnamefont {D.~K.~C.}\
  \bibnamefont {MacDonald}},\ }\href@noop {} {\emph {\bibinfo {title}
  {Thermoelectricity: An Introduction to the Principles}}}\ (\bibinfo
  {publisher} {Dover Publications},\ \bibinfo {year} {2006})\BibitemShut
  {NoStop}%
\bibitem [{\citenamefont {Hertz}\ and\ \citenamefont
  {Edwards}(1973)}]{Hertz1973}%
  \BibitemOpen
  \bibfield  {author} {\bibinfo {author} {\bibfnamefont {J.~A.}\ \bibnamefont
  {Hertz}}\ and\ \bibinfo {author} {\bibfnamefont {D.~M.}\ \bibnamefont
  {Edwards}},\ }\href {\doibase 10.1088/0305-4608/3/12/018} {\bibfield
  {journal} {\bibinfo  {journal} {Journal of Physics F: Metal Physics}\
  }\textbf {\bibinfo {volume} {3}},\ \bibinfo {pages} {2174} (\bibinfo {year}
  {1973})}\BibitemShut {NoStop}%
\bibitem [{\citenamefont {Edwards}\ and\ \citenamefont
  {Hertz}(1973)}]{Edwards1973}%
  \BibitemOpen
  \bibfield  {author} {\bibinfo {author} {\bibfnamefont {D.~M.}\ \bibnamefont
  {Edwards}}\ and\ \bibinfo {author} {\bibfnamefont {J.~A.}\ \bibnamefont
  {Hertz}},\ }\href {\doibase 10.1088/0305-4608/3/12/019} {\bibfield  {journal}
  {\bibinfo  {journal} {Journal of Physics F: Metal Physics}\ }\textbf
  {\bibinfo {volume} {3}},\ \bibinfo {pages} {2191} (\bibinfo {year}
  {1973})}\BibitemShut {NoStop}%
\bibitem [{\citenamefont {Lanzara}\ \emph {et~al.}(2001)\citenamefont
  {Lanzara}, \citenamefont {Bogdanov}, \citenamefont {Zhou}, \citenamefont
  {Kellar}, \citenamefont {Feng}, \citenamefont {Lu}, \citenamefont {Yoshida},
  \citenamefont {Eisaki}, \citenamefont {Fujimori}, \citenamefont {Kishio},
  \citenamefont {Shimoyama}, \citenamefont {Noda}, \citenamefont {Uchida},
  \citenamefont {Hussain},\ and\ \citenamefont {Shen}}]{Lanzara2001}%
  \BibitemOpen
  \bibfield  {author} {\bibinfo {author} {\bibfnamefont {A.}~\bibnamefont
  {Lanzara}}, \bibinfo {author} {\bibfnamefont {P.~V.}\ \bibnamefont
  {Bogdanov}}, \bibinfo {author} {\bibfnamefont {X.~J.}\ \bibnamefont {Zhou}},
  \bibinfo {author} {\bibfnamefont {S.~A.}\ \bibnamefont {Kellar}}, \bibinfo
  {author} {\bibfnamefont {D.~L.}\ \bibnamefont {Feng}}, \bibinfo {author}
  {\bibfnamefont {E.~D.}\ \bibnamefont {Lu}}, \bibinfo {author} {\bibfnamefont
  {T.}~\bibnamefont {Yoshida}}, \bibinfo {author} {\bibfnamefont
  {H.}~\bibnamefont {Eisaki}}, \bibinfo {author} {\bibfnamefont
  {A.}~\bibnamefont {Fujimori}}, \bibinfo {author} {\bibfnamefont
  {K.}~\bibnamefont {Kishio}}, \bibinfo {author} {\bibfnamefont {J.-I.}\
  \bibnamefont {Shimoyama}}, \bibinfo {author} {\bibfnamefont {T.}~\bibnamefont
  {Noda}}, \bibinfo {author} {\bibfnamefont {S.}~\bibnamefont {Uchida}},
  \bibinfo {author} {\bibfnamefont {Z.}~\bibnamefont {Hussain}}, \ and\
  \bibinfo {author} {\bibfnamefont {Z.-X.}\ \bibnamefont {Shen}},\ }\href
  {\doibase 10.1038/35087518} {\bibfield  {journal} {\bibinfo  {journal}
  {Nature}\ }\textbf {\bibinfo {volume} {412}},\ \bibinfo {pages} {510}
  (\bibinfo {year} {2001})}\BibitemShut {NoStop}%
\bibitem [{\citenamefont {Wang}\ \emph {et~al.}(2012)\citenamefont {Wang},
  \citenamefont {Pei}, \citenamefont {LaLonde},\ and\ \citenamefont
  {Snyder}}]{Wang2012}%
  \BibitemOpen
  \bibfield  {author} {\bibinfo {author} {\bibfnamefont {H.}~\bibnamefont
  {Wang}}, \bibinfo {author} {\bibfnamefont {Y.}~\bibnamefont {Pei}}, \bibinfo
  {author} {\bibfnamefont {A.~D.}\ \bibnamefont {LaLonde}}, \ and\ \bibinfo
  {author} {\bibfnamefont {G.~J.}\ \bibnamefont {Snyder}},\ }\href {\doibase
  10.1073/pnas.1111419109} {\bibfield  {journal} {\bibinfo  {journal} {PNAS}\
  }\textbf {\bibinfo {volume} {109}},\ \bibinfo {pages} {9705} (\bibinfo {year}
  {2012})}\BibitemShut {NoStop}%
\bibitem [{\citenamefont {van Elp}\ \emph {et~al.}(1991)\citenamefont {van
  Elp}, \citenamefont {Wieland}, \citenamefont {Eskes}, \citenamefont {Kuiper},
  \citenamefont {Sawatzky}, \citenamefont {de~Groot},\ and\ \citenamefont
  {Turner}}]{Elp1991}%
  \BibitemOpen
  \bibfield  {author} {\bibinfo {author} {\bibfnamefont {J.}~\bibnamefont {van
  Elp}}, \bibinfo {author} {\bibfnamefont {J.~L.}\ \bibnamefont {Wieland}},
  \bibinfo {author} {\bibfnamefont {H.}~\bibnamefont {Eskes}}, \bibinfo
  {author} {\bibfnamefont {P.}~\bibnamefont {Kuiper}}, \bibinfo {author}
  {\bibfnamefont {G.~A.}\ \bibnamefont {Sawatzky}}, \bibinfo {author}
  {\bibfnamefont {F.~M.~F.}\ \bibnamefont {de~Groot}}, \ and\ \bibinfo {author}
  {\bibfnamefont {T.~S.}\ \bibnamefont {Turner}},\ }\href {\doibase
  10.1103/PhysRevB.44.6090} {\bibfield  {journal} {\bibinfo  {journal} {Phys.
  Rev. B}\ }\textbf {\bibinfo {volume} {44}},\ \bibinfo {pages} {6090}
  (\bibinfo {year} {1991})}\BibitemShut {NoStop}%
\bibitem [{\citenamefont {Terasaki}\ \emph {et~al.}(1997)\citenamefont
  {Terasaki}, \citenamefont {Sasago},\ and\ \citenamefont
  {Uchinokura}}]{Terasaki1997}%
  \BibitemOpen
  \bibfield  {author} {\bibinfo {author} {\bibfnamefont {I.}~\bibnamefont
  {Terasaki}}, \bibinfo {author} {\bibfnamefont {Y.}~\bibnamefont {Sasago}}, \
  and\ \bibinfo {author} {\bibfnamefont {K.}~\bibnamefont {Uchinokura}},\
  }\href {\doibase 10.1103/PhysRevB.56.R12685} {\bibfield  {journal} {\bibinfo
  {journal} {Phys. Rev. B}\ }\textbf {\bibinfo {volume} {56}},\ \bibinfo
  {pages} {R12685} (\bibinfo {year} {1997})}\BibitemShut {NoStop}%
\bibitem [{\citenamefont {Sugiyama}\ \emph {et~al.}(2006)\citenamefont
  {Sugiyama}, \citenamefont {Nozaki}, \citenamefont {Ikedo}, \citenamefont
  {Mukai}, \citenamefont {Brewer}, \citenamefont {Ansaldo}, \citenamefont
  {Morris}, \citenamefont {Andreica}, \citenamefont {Amato}, \citenamefont
  {Fujii},\ and\ \citenamefont {Asamitsu}}]{Sugiyama2006}%
  \BibitemOpen
  \bibfield  {author} {\bibinfo {author} {\bibfnamefont {J.}~\bibnamefont
  {Sugiyama}}, \bibinfo {author} {\bibfnamefont {H.}~\bibnamefont {Nozaki}},
  \bibinfo {author} {\bibfnamefont {Y.}~\bibnamefont {Ikedo}}, \bibinfo
  {author} {\bibfnamefont {K.}~\bibnamefont {Mukai}}, \bibinfo {author}
  {\bibfnamefont {J.~H.}\ \bibnamefont {Brewer}}, \bibinfo {author}
  {\bibfnamefont {E.~J.}\ \bibnamefont {Ansaldo}}, \bibinfo {author}
  {\bibfnamefont {G.~D.}\ \bibnamefont {Morris}}, \bibinfo {author}
  {\bibfnamefont {D.}~\bibnamefont {Andreica}}, \bibinfo {author}
  {\bibfnamefont {A.}~\bibnamefont {Amato}}, \bibinfo {author} {\bibfnamefont
  {T.}~\bibnamefont {Fujii}}, \ and\ \bibinfo {author} {\bibfnamefont
  {A.}~\bibnamefont {Asamitsu}},\ }\href {\doibase
  10.1103/PhysRevLett.96.037206} {\bibfield  {journal} {\bibinfo  {journal}
  {Phys. Rev. Lett.}\ }\textbf {\bibinfo {volume} {96}},\ \bibinfo {pages}
  {037206} (\bibinfo {year} {2006})}\BibitemShut {NoStop}%
\bibitem [{\citenamefont {Shibasaki}\ \emph {et~al.}(2010)\citenamefont
  {Shibasaki}, \citenamefont {Nakano}, \citenamefont {Terasaki}, \citenamefont
  {Yubuta},\ and\ \citenamefont {Kajitani}}]{Shibasaki_2010}%
  \BibitemOpen
  \bibfield  {author} {\bibinfo {author} {\bibfnamefont {S.}~\bibnamefont
  {Shibasaki}}, \bibinfo {author} {\bibfnamefont {T.}~\bibnamefont {Nakano}},
  \bibinfo {author} {\bibfnamefont {I.}~\bibnamefont {Terasaki}}, \bibinfo
  {author} {\bibfnamefont {K.}~\bibnamefont {Yubuta}}, \ and\ \bibinfo {author}
  {\bibfnamefont {T.}~\bibnamefont {Kajitani}},\ }\href {\doibase
  10.1088/0953-8984/22/11/115603} {\bibfield  {journal} {\bibinfo  {journal}
  {Journal of Physics: Condensed Matter}\ }\textbf {\bibinfo {volume} {22}},\
  \bibinfo {pages} {115603} (\bibinfo {year} {2010})}\BibitemShut {NoStop}%
\bibitem [{\citenamefont {Schaak}\ \emph {et~al.}(2003)\citenamefont {Schaak},
  \citenamefont {Klimczuk}, \citenamefont {Foo},\ and\ \citenamefont
  {Cava}}]{Schaak2003}%
  \BibitemOpen
  \bibfield  {author} {\bibinfo {author} {\bibfnamefont {R.}~\bibnamefont
  {Schaak}}, \bibinfo {author} {\bibfnamefont {T.}~\bibnamefont {Klimczuk}},
  \bibinfo {author} {\bibfnamefont {M.}~\bibnamefont {Foo}}, \ and\ \bibinfo
  {author} {\bibfnamefont {R.}~\bibnamefont {Cava}},\ }\href {\doibase
  10.1038/nature01877} {\bibfield  {journal} {\bibinfo  {journal} {Nature}\
  }\textbf {\bibinfo {volume} {424}},\ \bibinfo {pages} {527} (\bibinfo {year}
  {2003})}\BibitemShut {NoStop}%
\bibitem [{\citenamefont {Lee}\ \emph {et~al.}(2006)\citenamefont {Lee},
  \citenamefont {Viciu}, \citenamefont {Li}, \citenamefont {Wang},
  \citenamefont {Foo}, \citenamefont {Watauchi}, \citenamefont {Pascal},
  \citenamefont {Cava},\ and\ \citenamefont {Ong}}]{Lee2006}%
  \BibitemOpen
  \bibfield  {author} {\bibinfo {author} {\bibfnamefont {M.}~\bibnamefont
  {Lee}}, \bibinfo {author} {\bibfnamefont {L.}~\bibnamefont {Viciu}}, \bibinfo
  {author} {\bibfnamefont {L.}~\bibnamefont {Li}}, \bibinfo {author}
  {\bibfnamefont {Y.}~\bibnamefont {Wang}}, \bibinfo {author} {\bibfnamefont
  {M.}~\bibnamefont {Foo}}, \bibinfo {author} {\bibfnamefont {S.}~\bibnamefont
  {Watauchi}}, \bibinfo {author} {\bibfnamefont {R.}~\bibnamefont {Pascal}},
  \bibinfo {author} {\bibfnamefont {R.}~\bibnamefont {Cava}}, \ and\ \bibinfo
  {author} {\bibfnamefont {N.}~\bibnamefont {Ong}},\ }\href {\doibase
  10.1038/nmat1669} {\bibfield  {journal} {\bibinfo  {journal} {Nature
  materials}\ }\textbf {\bibinfo {volume} {5}},\ \bibinfo {pages} {537}
  (\bibinfo {year} {2006})}\BibitemShut {NoStop}%
\bibitem [{\citenamefont {Helme}\ \emph {et~al.}(2005)\citenamefont {Helme},
  \citenamefont {Boothroyd}, \citenamefont {Coldea}, \citenamefont
  {Prabhakaran}, \citenamefont {Tennant}, \citenamefont {Hiess},\ and\
  \citenamefont {Kulda}}]{Helme2005}%
  \BibitemOpen
  \bibfield  {author} {\bibinfo {author} {\bibfnamefont {L.~M.}\ \bibnamefont
  {Helme}}, \bibinfo {author} {\bibfnamefont {A.~T.}\ \bibnamefont
  {Boothroyd}}, \bibinfo {author} {\bibfnamefont {R.}~\bibnamefont {Coldea}},
  \bibinfo {author} {\bibfnamefont {D.}~\bibnamefont {Prabhakaran}}, \bibinfo
  {author} {\bibfnamefont {D.~A.}\ \bibnamefont {Tennant}}, \bibinfo {author}
  {\bibfnamefont {A.}~\bibnamefont {Hiess}}, \ and\ \bibinfo {author}
  {\bibfnamefont {J.}~\bibnamefont {Kulda}},\ }\href {\doibase
  10.1103/PhysRevLett.94.157206} {\bibfield  {journal} {\bibinfo  {journal}
  {Phys. Rev. Lett.}\ }\textbf {\bibinfo {volume} {94}},\ \bibinfo {pages}
  {157206} (\bibinfo {year} {2005})}\BibitemShut {NoStop}%
\bibitem [{\citenamefont {Bayrakci}\ \emph {et~al.}(2005)\citenamefont
  {Bayrakci}, \citenamefont {Mirebeau}, \citenamefont {Bourges}, \citenamefont
  {Sidis}, \citenamefont {Enderle}, \citenamefont {Mesot}, \citenamefont
  {Chen}, \citenamefont {Lin},\ and\ \citenamefont {Keimer}}]{Bayrakci2005}%
  \BibitemOpen
  \bibfield  {author} {\bibinfo {author} {\bibfnamefont {S.~P.}\ \bibnamefont
  {Bayrakci}}, \bibinfo {author} {\bibfnamefont {I.}~\bibnamefont {Mirebeau}},
  \bibinfo {author} {\bibfnamefont {P.}~\bibnamefont {Bourges}}, \bibinfo
  {author} {\bibfnamefont {Y.}~\bibnamefont {Sidis}}, \bibinfo {author}
  {\bibfnamefont {M.}~\bibnamefont {Enderle}}, \bibinfo {author} {\bibfnamefont
  {J.}~\bibnamefont {Mesot}}, \bibinfo {author} {\bibfnamefont {D.~P.}\
  \bibnamefont {Chen}}, \bibinfo {author} {\bibfnamefont {C.~T.}\ \bibnamefont
  {Lin}}, \ and\ \bibinfo {author} {\bibfnamefont {B.}~\bibnamefont {Keimer}},\
  }\href {\doibase 10.1103/PhysRevLett.94.157205} {\bibfield  {journal}
  {\bibinfo  {journal} {Phys. Rev. Lett.}\ }\textbf {\bibinfo {volume} {94}},\
  \bibinfo {pages} {157205} (\bibinfo {year} {2005})}\BibitemShut {NoStop}%
\bibitem [{\citenamefont {Foo}\ \emph {et~al.}(2004)\citenamefont {Foo},
  \citenamefont {Wang}, \citenamefont {Watauchi}, \citenamefont {Zandbergen},
  \citenamefont {He}, \citenamefont {Cava},\ and\ \citenamefont
  {Ong}}]{Foo2004}%
  \BibitemOpen
  \bibfield  {author} {\bibinfo {author} {\bibfnamefont {M.~L.}\ \bibnamefont
  {Foo}}, \bibinfo {author} {\bibfnamefont {Y.}~\bibnamefont {Wang}}, \bibinfo
  {author} {\bibfnamefont {S.}~\bibnamefont {Watauchi}}, \bibinfo {author}
  {\bibfnamefont {H.~W.}\ \bibnamefont {Zandbergen}}, \bibinfo {author}
  {\bibfnamefont {T.}~\bibnamefont {He}}, \bibinfo {author} {\bibfnamefont
  {R.~J.}\ \bibnamefont {Cava}}, \ and\ \bibinfo {author} {\bibfnamefont
  {N.~P.}\ \bibnamefont {Ong}},\ }\href {\doibase
  10.1103/PhysRevLett.92.247001} {\bibfield  {journal} {\bibinfo  {journal}
  {Phys. Rev. Lett.}\ }\textbf {\bibinfo {volume} {92}},\ \bibinfo {pages}
  {247001} (\bibinfo {year} {2004})}\BibitemShut {NoStop}%
\bibitem [{\citenamefont {Okazaki}\ \emph {et~al.}(2011)\citenamefont
  {Okazaki}, \citenamefont {Nishina}, \citenamefont {Yasui}, \citenamefont
  {Shibasaki},\ and\ \citenamefont {Terasaki}}]{Okazaki2011}%
  \BibitemOpen
  \bibfield  {author} {\bibinfo {author} {\bibfnamefont {R.}~\bibnamefont
  {Okazaki}}, \bibinfo {author} {\bibfnamefont {Y.}~\bibnamefont {Nishina}},
  \bibinfo {author} {\bibfnamefont {Y.}~\bibnamefont {Yasui}}, \bibinfo
  {author} {\bibfnamefont {S.}~\bibnamefont {Shibasaki}}, \ and\ \bibinfo
  {author} {\bibfnamefont {I.}~\bibnamefont {Terasaki}},\ }\href {\doibase
  10.1103/PhysRevB.84.075110} {\bibfield  {journal} {\bibinfo  {journal} {Phys.
  Rev. B}\ }\textbf {\bibinfo {volume} {84}},\ \bibinfo {pages} {075110}
  (\bibinfo {year} {2011})}\BibitemShut {NoStop}%
\bibitem [{\citenamefont {Koshibae}\ and\ \citenamefont
  {Maekawa}(2001)}]{Koshibae2001}%
  \BibitemOpen
  \bibfield  {author} {\bibinfo {author} {\bibfnamefont {W.}~\bibnamefont
  {Koshibae}}\ and\ \bibinfo {author} {\bibfnamefont {S.}~\bibnamefont
  {Maekawa}},\ }\href {\doibase 10.1103/PhysRevLett.87.236603} {\bibfield
  {journal} {\bibinfo  {journal} {Phys. Rev. Lett.}\ }\textbf {\bibinfo
  {volume} {87}},\ \bibinfo {pages} {236603} (\bibinfo {year}
  {2001})}\BibitemShut {NoStop}%
\bibitem [{\citenamefont {Wang}\ \emph {et~al.}(2003)\citenamefont {Wang},
  \citenamefont {Rogado}, \citenamefont {Cava},\ and\ \citenamefont
  {Ong}}]{Wang2003}%
  \BibitemOpen
  \bibfield  {author} {\bibinfo {author} {\bibfnamefont {Y.}~\bibnamefont
  {Wang}}, \bibinfo {author} {\bibfnamefont {N.}~\bibnamefont {Rogado}},
  \bibinfo {author} {\bibfnamefont {R.}~\bibnamefont {Cava}}, \ and\ \bibinfo
  {author} {\bibfnamefont {N.}~\bibnamefont {Ong}},\ }\href {\doibase
  10.1038/nature01639} {\bibfield  {journal} {\bibinfo  {journal} {Nature}\
  }\textbf {\bibinfo {volume} {423}},\ \bibinfo {pages} {425} (\bibinfo {year}
  {2003})}\BibitemShut {NoStop}%
\bibitem [{\citenamefont {Donkov}\ \emph {et~al.}(2008)\citenamefont {Donkov},
  \citenamefont {Korshunov}, \citenamefont {Eremin}, \citenamefont {Lemmens},
  \citenamefont {Gnezdilov}, \citenamefont {Chou},\ and\ \citenamefont
  {Lin}}]{Donkov2008}%
  \BibitemOpen
  \bibfield  {author} {\bibinfo {author} {\bibfnamefont {A.}~\bibnamefont
  {Donkov}}, \bibinfo {author} {\bibfnamefont {M.~M.}\ \bibnamefont
  {Korshunov}}, \bibinfo {author} {\bibfnamefont {I.}~\bibnamefont {Eremin}},
  \bibinfo {author} {\bibfnamefont {P.}~\bibnamefont {Lemmens}}, \bibinfo
  {author} {\bibfnamefont {V.}~\bibnamefont {Gnezdilov}}, \bibinfo {author}
  {\bibfnamefont {F.~C.}\ \bibnamefont {Chou}}, \ and\ \bibinfo {author}
  {\bibfnamefont {C.~T.}\ \bibnamefont {Lin}},\ }\href {\doibase
  10.1103/PhysRevB.77.100504} {\bibfield  {journal} {\bibinfo  {journal} {Phys.
  Rev. B}\ }\textbf {\bibinfo {volume} {77}},\ \bibinfo {pages} {100504}
  (\bibinfo {year} {2008})}\BibitemShut {NoStop}%
\bibitem [{\citenamefont {Yao}\ \emph {et~al.}(2012)\citenamefont {Yao},
  \citenamefont {Zhang}, \citenamefont {Zhou}, \citenamefont {Chen},
  \citenamefont {Zhang}, \citenamefont {Gu}, \citenamefont {Dong},\ and\
  \citenamefont {Chen}}]{Yao2012}%
  \BibitemOpen
  \bibfield  {author} {\bibinfo {author} {\bibfnamefont {S.~H.}\ \bibnamefont
  {Yao}}, \bibinfo {author} {\bibfnamefont {B.~B.}\ \bibnamefont {Zhang}},
  \bibinfo {author} {\bibfnamefont {J.}~\bibnamefont {Zhou}}, \bibinfo {author}
  {\bibfnamefont {Y.~B.}\ \bibnamefont {Chen}}, \bibinfo {author}
  {\bibfnamefont {S.~T.}\ \bibnamefont {Zhang}}, \bibinfo {author}
  {\bibfnamefont {Z.~B.}\ \bibnamefont {Gu}}, \bibinfo {author} {\bibfnamefont
  {S.~T.}\ \bibnamefont {Dong}}, \ and\ \bibinfo {author} {\bibfnamefont
  {Y.~F.}\ \bibnamefont {Chen}},\ }\href {\doibase 10.1063/1.4767464}
  {\bibfield  {journal} {\bibinfo  {journal} {AIP Adv.}\ }\textbf {\bibinfo
  {volume} {2}},\ \bibinfo {pages} {042140} (\bibinfo {year}
  {2012})}\BibitemShut {NoStop}%
\bibitem [{\citenamefont {Saeed}\ \emph {et~al.}(2012)\citenamefont {Saeed},
  \citenamefont {Singh},\ and\ \citenamefont {Schwingenschlögl}}]{Saeed2012}%
  \BibitemOpen
  \bibfield  {author} {\bibinfo {author} {\bibfnamefont {Y.}~\bibnamefont
  {Saeed}}, \bibinfo {author} {\bibfnamefont {N.}~\bibnamefont {Singh}}, \ and\
  \bibinfo {author} {\bibfnamefont {U.}~\bibnamefont {Schwingenschlögl}},\
  }\href {\doibase 10.1002/adfm.201103106} {\bibfield  {journal} {\bibinfo
  {journal} {Adv. Funct. Mater.}\ }\textbf {\bibinfo {volume} {22}},\ \bibinfo
  {pages} {2792} (\bibinfo {year} {2012})}\BibitemShut {NoStop}%
\bibitem [{\citenamefont {Hasan}\ \emph {et~al.}(2004)\citenamefont {Hasan},
  \citenamefont {Chuang}, \citenamefont {Qian}, \citenamefont {Li},
  \citenamefont {Kong}, \citenamefont {Kuprin}, \citenamefont {Fedorov},
  \citenamefont {Kimmerling}, \citenamefont {Rotenberg}, \citenamefont
  {Rossnagel}, \citenamefont {Hussain}, \citenamefont {Koh}, \citenamefont
  {Rogado}, \citenamefont {Foo},\ and\ \citenamefont {Cava}}]{Hasan2004}%
  \BibitemOpen
  \bibfield  {author} {\bibinfo {author} {\bibfnamefont {M.~Z.}\ \bibnamefont
  {Hasan}}, \bibinfo {author} {\bibfnamefont {Y.-D.}\ \bibnamefont {Chuang}},
  \bibinfo {author} {\bibfnamefont {D.}~\bibnamefont {Qian}}, \bibinfo {author}
  {\bibfnamefont {Y.~W.}\ \bibnamefont {Li}}, \bibinfo {author} {\bibfnamefont
  {Y.}~\bibnamefont {Kong}}, \bibinfo {author} {\bibfnamefont {A.}~\bibnamefont
  {Kuprin}}, \bibinfo {author} {\bibfnamefont {A.~V.}\ \bibnamefont {Fedorov}},
  \bibinfo {author} {\bibfnamefont {R.}~\bibnamefont {Kimmerling}}, \bibinfo
  {author} {\bibfnamefont {E.}~\bibnamefont {Rotenberg}}, \bibinfo {author}
  {\bibfnamefont {K.}~\bibnamefont {Rossnagel}}, \bibinfo {author}
  {\bibfnamefont {Z.}~\bibnamefont {Hussain}}, \bibinfo {author} {\bibfnamefont
  {H.}~\bibnamefont {Koh}}, \bibinfo {author} {\bibfnamefont {N.~S.}\
  \bibnamefont {Rogado}}, \bibinfo {author} {\bibfnamefont {M.~L.}\
  \bibnamefont {Foo}}, \ and\ \bibinfo {author} {\bibfnamefont {R.~J.}\
  \bibnamefont {Cava}},\ }\href {\doibase 10.1103/PhysRevLett.92.246402}
  {\bibfield  {journal} {\bibinfo  {journal} {Phys. Rev. Lett.}\ }\textbf
  {\bibinfo {volume} {92}},\ \bibinfo {pages} {246402} (\bibinfo {year}
  {2004})}\BibitemShut {NoStop}%
\bibitem [{\citenamefont {Yang}\ \emph {et~al.}(2005)\citenamefont {Yang},
  \citenamefont {Pan}, \citenamefont {Sekharan}, \citenamefont {Sato},
  \citenamefont {Souma}, \citenamefont {Takahashi}, \citenamefont {Jin},
  \citenamefont {Sales}, \citenamefont {Mandrus}, \citenamefont {Fedorov},
  \citenamefont {Wang},\ and\ \citenamefont {Ding}}]{Yang2005}%
  \BibitemOpen
  \bibfield  {author} {\bibinfo {author} {\bibfnamefont {H.-B.}\ \bibnamefont
  {Yang}}, \bibinfo {author} {\bibfnamefont {Z.-H.}\ \bibnamefont {Pan}},
  \bibinfo {author} {\bibfnamefont {A.~K.~P.}\ \bibnamefont {Sekharan}},
  \bibinfo {author} {\bibfnamefont {T.}~\bibnamefont {Sato}}, \bibinfo {author}
  {\bibfnamefont {S.}~\bibnamefont {Souma}}, \bibinfo {author} {\bibfnamefont
  {T.}~\bibnamefont {Takahashi}}, \bibinfo {author} {\bibfnamefont
  {R.}~\bibnamefont {Jin}}, \bibinfo {author} {\bibfnamefont {B.~C.}\
  \bibnamefont {Sales}}, \bibinfo {author} {\bibfnamefont {D.}~\bibnamefont
  {Mandrus}}, \bibinfo {author} {\bibfnamefont {A.~V.}\ \bibnamefont
  {Fedorov}}, \bibinfo {author} {\bibfnamefont {Z.}~\bibnamefont {Wang}}, \
  and\ \bibinfo {author} {\bibfnamefont {H.}~\bibnamefont {Ding}},\ }\href
  {\doibase 10.1103/PhysRevLett.95.146401} {\bibfield  {journal} {\bibinfo
  {journal} {Phys. Rev. Lett.}\ }\textbf {\bibinfo {volume} {95}},\ \bibinfo
  {pages} {146401} (\bibinfo {year} {2005})}\BibitemShut {NoStop}%
\bibitem [{\citenamefont {Chen}\ \emph {et~al.}(2017)\citenamefont {Chen},
  \citenamefont {He}, \citenamefont {Zong}, \citenamefont {Zhang},
  \citenamefont {Hashimoto}, \citenamefont {Zhang}, \citenamefont {Yao},
  \citenamefont {Chen}, \citenamefont {Zhou}, \citenamefont {Chen},
  \citenamefont {Mo}, \citenamefont {Hussain}, \citenamefont {Lu},\ and\
  \citenamefont {Shen}}]{Chen2017}%
  \BibitemOpen
  \bibfield  {author} {\bibinfo {author} {\bibfnamefont {S.-D.}\ \bibnamefont
  {Chen}}, \bibinfo {author} {\bibfnamefont {Y.}~\bibnamefont {He}}, \bibinfo
  {author} {\bibfnamefont {A.}~\bibnamefont {Zong}}, \bibinfo {author}
  {\bibfnamefont {Y.}~\bibnamefont {Zhang}}, \bibinfo {author} {\bibfnamefont
  {M.}~\bibnamefont {Hashimoto}}, \bibinfo {author} {\bibfnamefont {B.-B.}\
  \bibnamefont {Zhang}}, \bibinfo {author} {\bibfnamefont {S.-H.}\ \bibnamefont
  {Yao}}, \bibinfo {author} {\bibfnamefont {Y.-B.}\ \bibnamefont {Chen}},
  \bibinfo {author} {\bibfnamefont {J.}~\bibnamefont {Zhou}}, \bibinfo {author}
  {\bibfnamefont {Y.-F.}\ \bibnamefont {Chen}}, \bibinfo {author}
  {\bibfnamefont {S.-K.}\ \bibnamefont {Mo}}, \bibinfo {author} {\bibfnamefont
  {Z.}~\bibnamefont {Hussain}}, \bibinfo {author} {\bibfnamefont
  {D.}~\bibnamefont {Lu}}, \ and\ \bibinfo {author} {\bibfnamefont {Z.-X.}\
  \bibnamefont {Shen}},\ }\href
  {https://link.aps.org/doi/10.1103/PhysRevB.96.081109} {\bibfield  {journal}
  {\bibinfo  {journal} {Phys. Rev. B}\ }\textbf {\bibinfo {volume} {96}},\
  \bibinfo {pages} {081109} (\bibinfo {year} {2017})}\BibitemShut {NoStop}%
\bibitem [{\citenamefont {Singh}(2000)}]{Singh2000}%
  \BibitemOpen
  \bibfield  {author} {\bibinfo {author} {\bibfnamefont {D.~J.}\ \bibnamefont
  {Singh}},\ }\href {\doibase 10.1103/PhysRevB.61.13397} {\bibfield  {journal}
  {\bibinfo  {journal} {Phys. Rev. B}\ }\textbf {\bibinfo {volume} {61}},\
  \bibinfo {pages} {13397} (\bibinfo {year} {2000})}\BibitemShut {NoStop}%
\bibitem [{\citenamefont {Arakane}\ \emph {et~al.}(2011)\citenamefont
  {Arakane}, \citenamefont {Sato}, \citenamefont {Takahashi}, \citenamefont
  {Fujii},\ and\ \citenamefont {Asamitsu}}]{Arakane2011}%
  \BibitemOpen
  \bibfield  {author} {\bibinfo {author} {\bibfnamefont {T.}~\bibnamefont
  {Arakane}}, \bibinfo {author} {\bibfnamefont {T.}~\bibnamefont {Sato}},
  \bibinfo {author} {\bibfnamefont {T.}~\bibnamefont {Takahashi}}, \bibinfo
  {author} {\bibfnamefont {T.}~\bibnamefont {Fujii}}, \ and\ \bibinfo {author}
  {\bibfnamefont {A.}~\bibnamefont {Asamitsu}},\ }\href {\doibase
  10.1088/1367-2630/13/4/043021} {\bibfield  {journal} {\bibinfo  {journal}
  {New J. Phys.}\ }\textbf {\bibinfo {volume} {13}},\ \bibinfo {pages} {043021}
  (\bibinfo {year} {2011})}\BibitemShut {NoStop}%
\bibitem [{\citenamefont {Okamoto}\ \emph {et~al.}(2017)\citenamefont
  {Okamoto}, \citenamefont {Matsumoto}, \citenamefont {Yagihara}, \citenamefont
  {Iwai}, \citenamefont {Miyoshi}, \citenamefont {Takeuchi}, \citenamefont
  {Horiba}, \citenamefont {Kobayashi}, \citenamefont {Ono}, \citenamefont
  {Kumigashira}, \citenamefont {Saini},\ and\ \citenamefont
  {Mizokawa}}]{Okamoto2017}%
  \BibitemOpen
  \bibfield  {author} {\bibinfo {author} {\bibfnamefont {Y.}~\bibnamefont
  {Okamoto}}, \bibinfo {author} {\bibfnamefont {R.}~\bibnamefont {Matsumoto}},
  \bibinfo {author} {\bibfnamefont {T.}~\bibnamefont {Yagihara}}, \bibinfo
  {author} {\bibfnamefont {C.}~\bibnamefont {Iwai}}, \bibinfo {author}
  {\bibfnamefont {K.}~\bibnamefont {Miyoshi}}, \bibinfo {author} {\bibfnamefont
  {J.}~\bibnamefont {Takeuchi}}, \bibinfo {author} {\bibfnamefont
  {K.}~\bibnamefont {Horiba}}, \bibinfo {author} {\bibfnamefont
  {M.}~\bibnamefont {Kobayashi}}, \bibinfo {author} {\bibfnamefont
  {K.}~\bibnamefont {Ono}}, \bibinfo {author} {\bibfnamefont {H.}~\bibnamefont
  {Kumigashira}}, \bibinfo {author} {\bibfnamefont {N.~L.}\ \bibnamefont
  {Saini}}, \ and\ \bibinfo {author} {\bibfnamefont {T.}~\bibnamefont
  {Mizokawa}},\ }\href {\doibase 10.1103/PhysRevB.96.125147} {\bibfield
  {journal} {\bibinfo  {journal} {Phys. Rev. B}\ }\textbf {\bibinfo {volume}
  {96}},\ \bibinfo {pages} {125147} (\bibinfo {year} {2017})}\BibitemShut
  {NoStop}%
\bibitem [{\citenamefont {Reinert}\ and\ \citenamefont {Hüfner}()}]{Reinert}%
  \BibitemOpen
  \bibfield  {author} {\bibinfo {author} {\bibfnamefont {F.}~\bibnamefont
  {Reinert}}\ and\ \bibinfo {author} {\bibfnamefont {S.}~\bibnamefont
  {Hüfner}},\ }in\ \href {\doibase 10.1007/3-540-68133-7_2} {\emph {\bibinfo
  {booktitle} {Very High Resolution Photoelectron Spectroscopy}}}\ (\bibinfo
  {publisher} {Springer Berlin Heidelberg})\ pp.\ \bibinfo {pages}
  {13--53}\BibitemShut {NoStop}%
\bibitem [{\citenamefont {Nicolaou}\ \emph {et~al.}(2010)\citenamefont
  {Nicolaou}, \citenamefont {Brouet}, \citenamefont {Zacchigna}, \citenamefont
  {Vobornik}, \citenamefont {Tejeda}, \citenamefont {Taleb-Ibrahimi},
  \citenamefont {Le~F\`evre}, \citenamefont {Bertran}, \citenamefont
  {H\'ebert}, \citenamefont {Muguerra},\ and\ \citenamefont
  {Grebille}}]{Nicolaou2010}%
  \BibitemOpen
  \bibfield  {author} {\bibinfo {author} {\bibfnamefont {A.}~\bibnamefont
  {Nicolaou}}, \bibinfo {author} {\bibfnamefont {V.}~\bibnamefont {Brouet}},
  \bibinfo {author} {\bibfnamefont {M.}~\bibnamefont {Zacchigna}}, \bibinfo
  {author} {\bibfnamefont {I.}~\bibnamefont {Vobornik}}, \bibinfo {author}
  {\bibfnamefont {A.}~\bibnamefont {Tejeda}}, \bibinfo {author} {\bibfnamefont
  {A.}~\bibnamefont {Taleb-Ibrahimi}}, \bibinfo {author} {\bibfnamefont
  {P.}~\bibnamefont {Le~F\`evre}}, \bibinfo {author} {\bibfnamefont
  {F.}~\bibnamefont {Bertran}}, \bibinfo {author} {\bibfnamefont
  {S.}~\bibnamefont {H\'ebert}}, \bibinfo {author} {\bibfnamefont
  {H.}~\bibnamefont {Muguerra}}, \ and\ \bibinfo {author} {\bibfnamefont
  {D.}~\bibnamefont {Grebille}},\ }\href {\doibase
  10.1103/PhysRevLett.104.056403} {\bibfield  {journal} {\bibinfo  {journal}
  {Phys. Rev. Lett.}\ }\textbf {\bibinfo {volume} {104}},\ \bibinfo {pages}
  {056403} (\bibinfo {year} {2010})}\BibitemShut {NoStop}%
\bibitem [{\citenamefont {Fink}\ \emph {et~al.}(2007)\citenamefont {Fink},
  \citenamefont {Borisenko}, \citenamefont {Kordyuk}, \citenamefont {Koitzsch},
  \citenamefont {Geck}, \citenamefont {Zabolotnyy}, \citenamefont {Knupfer},
  \citenamefont {Büchner},\ and\ \citenamefont {Berger}}]{Fink}%
  \BibitemOpen
  \bibfield  {author} {\bibinfo {author} {\bibfnamefont {J.}~\bibnamefont
  {Fink}}, \bibinfo {author} {\bibfnamefont {S.}~\bibnamefont {Borisenko}},
  \bibinfo {author} {\bibfnamefont {A.}~\bibnamefont {Kordyuk}}, \bibinfo
  {author} {\bibfnamefont {A.}~\bibnamefont {Koitzsch}}, \bibinfo {author}
  {\bibfnamefont {J.}~\bibnamefont {Geck}}, \bibinfo {author} {\bibfnamefont
  {V.}~\bibnamefont {Zabolotnyy}}, \bibinfo {author} {\bibfnamefont
  {M.}~\bibnamefont {Knupfer}}, \bibinfo {author} {\bibfnamefont
  {B.}~\bibnamefont {Büchner}}, \ and\ \bibinfo {author} {\bibfnamefont
  {H.}~\bibnamefont {Berger}},\ }in\ \href {\doibase 10.1007/3-540-68133-7_11}
  {\emph {\bibinfo {booktitle} {Very High Resolution Photoelectron
  Spectroscopy}}},\ \bibinfo {editor} {edited by\ \bibinfo {editor}
  {\bibfnamefont {H.~S.}\ \bibnamefont {(eds)}}}\ (\bibinfo  {publisher}
  {Springer Berlin Heidelberg},\ \bibinfo {year} {2007})\ pp.\ \bibinfo {pages}
  {295--325}\BibitemShut {NoStop}%
\bibitem [{\citenamefont {Geck}\ \emph {et~al.}(2007)\citenamefont {Geck},
  \citenamefont {Borisenko}, \citenamefont {Berger}, \citenamefont {Eschrig},
  \citenamefont {Fink}, \citenamefont {Knupfer}, \citenamefont {Koepernik},
  \citenamefont {Koitzsch}, \citenamefont {Kordyuk}, \citenamefont
  {Zabolotnyy},\ and\ \citenamefont {B\"uchner}}]{Geck2007}%
  \BibitemOpen
  \bibfield  {author} {\bibinfo {author} {\bibfnamefont {J.}~\bibnamefont
  {Geck}}, \bibinfo {author} {\bibfnamefont {S.~V.}\ \bibnamefont {Borisenko}},
  \bibinfo {author} {\bibfnamefont {H.}~\bibnamefont {Berger}}, \bibinfo
  {author} {\bibfnamefont {H.}~\bibnamefont {Eschrig}}, \bibinfo {author}
  {\bibfnamefont {J.}~\bibnamefont {Fink}}, \bibinfo {author} {\bibfnamefont
  {M.}~\bibnamefont {Knupfer}}, \bibinfo {author} {\bibfnamefont
  {K.}~\bibnamefont {Koepernik}}, \bibinfo {author} {\bibfnamefont
  {A.}~\bibnamefont {Koitzsch}}, \bibinfo {author} {\bibfnamefont {A.~A.}\
  \bibnamefont {Kordyuk}}, \bibinfo {author} {\bibfnamefont {V.~B.}\
  \bibnamefont {Zabolotnyy}}, \ and\ \bibinfo {author} {\bibfnamefont
  {B.}~\bibnamefont {B\"uchner}},\ }\href {\doibase
  10.1103/PhysRevLett.99.046403} {\bibfield  {journal} {\bibinfo  {journal}
  {Phys. Rev. Lett.}\ }\textbf {\bibinfo {volume} {99}},\ \bibinfo {pages}
  {046403} (\bibinfo {year} {2007})}\BibitemShut {NoStop}%
\bibitem [{\citenamefont {Luttinger}(1960)}]{Luttinger1960}%
  \BibitemOpen
  \bibfield  {author} {\bibinfo {author} {\bibfnamefont {J.~M.}\ \bibnamefont
  {Luttinger}},\ }\href {\doibase 10.1103/PhysRev.119.1153} {\bibfield
  {journal} {\bibinfo  {journal} {Phys. Rev.}\ }\textbf {\bibinfo {volume}
  {119}},\ \bibinfo {pages} {1153} (\bibinfo {year} {1960})}\BibitemShut
  {NoStop}%
\bibitem [{\citenamefont {Zhang}\ \emph {et~al.}(2015)\citenamefont {Zhang},
  \citenamefont {Zhang}, \citenamefont {Dong}, \citenamefont {Lv},
  \citenamefont {Chen}, \citenamefont {Yao}, \citenamefont {Zhang},
  \citenamefont {Gu}, \citenamefont {Zhou}, \citenamefont {Guedes},
  \citenamefont {Yu},\ and\ \citenamefont {Chen}}]{Zhang2015}%
  \BibitemOpen
  \bibfield  {author} {\bibinfo {author} {\bibfnamefont {B.-B.}\ \bibnamefont
  {Zhang}}, \bibinfo {author} {\bibfnamefont {N.}~\bibnamefont {Zhang}},
  \bibinfo {author} {\bibfnamefont {S.-T.}\ \bibnamefont {Dong}}, \bibinfo
  {author} {\bibfnamefont {Y.}~\bibnamefont {Lv}}, \bibinfo {author}
  {\bibfnamefont {Y.~B.}\ \bibnamefont {Chen}}, \bibinfo {author}
  {\bibfnamefont {S.}~\bibnamefont {Yao}}, \bibinfo {author} {\bibfnamefont
  {S.-T.}\ \bibnamefont {Zhang}}, \bibinfo {author} {\bibfnamefont {Z.-B.}\
  \bibnamefont {Gu}}, \bibinfo {author} {\bibfnamefont {J.}~\bibnamefont
  {Zhou}}, \bibinfo {author} {\bibfnamefont {I.}~\bibnamefont {Guedes}},
  \bibinfo {author} {\bibfnamefont {D.}~\bibnamefont {Yu}}, \ and\ \bibinfo
  {author} {\bibfnamefont {Y.-F.}\ \bibnamefont {Chen}},\ }\href {\doibase
  10.1063/1.4928384} {\bibfield  {journal} {\bibinfo  {journal} {{AIP}
  Advances}\ }\textbf {\bibinfo {volume} {5}},\ \bibinfo {pages} {087111}
  (\bibinfo {year} {2015})}\BibitemShut {NoStop}%
\bibitem [{\citenamefont {Sch\"afer}\ \emph {et~al.}(2004)\citenamefont
  {Sch\"afer}, \citenamefont {Schrupp}, \citenamefont {Rotenberg},
  \citenamefont {Rossnagel}, \citenamefont {Koh}, \citenamefont {Blaha},\ and\
  \citenamefont {Claessen}}]{Schaefer2004}%
  \BibitemOpen
  \bibfield  {author} {\bibinfo {author} {\bibfnamefont {J.}~\bibnamefont
  {Sch\"afer}}, \bibinfo {author} {\bibfnamefont {D.}~\bibnamefont {Schrupp}},
  \bibinfo {author} {\bibfnamefont {E.}~\bibnamefont {Rotenberg}}, \bibinfo
  {author} {\bibfnamefont {K.}~\bibnamefont {Rossnagel}}, \bibinfo {author}
  {\bibfnamefont {H.}~\bibnamefont {Koh}}, \bibinfo {author} {\bibfnamefont
  {P.}~\bibnamefont {Blaha}}, \ and\ \bibinfo {author} {\bibfnamefont
  {R.}~\bibnamefont {Claessen}},\ }\href {\doibase
  10.1103/PhysRevLett.92.097205} {\bibfield  {journal} {\bibinfo  {journal}
  {Phys. Rev. Lett.}\ }\textbf {\bibinfo {volume} {92}},\ \bibinfo {pages}
  {097205} (\bibinfo {year} {2004})}\BibitemShut {NoStop}%
\bibitem [{\citenamefont {Mazzola}\ \emph {et~al.}(2013)\citenamefont
  {Mazzola}, \citenamefont {Wells}, \citenamefont {Yakimova}, \citenamefont
  {Ulstrup}, \citenamefont {Miwa}, \citenamefont {Balog}, \citenamefont
  {Bianchi}, \citenamefont {Leandersson}, \citenamefont {Adell}, \citenamefont
  {Hofmann},\ and\ \citenamefont {Balasubramanian}}]{Mazzola2013}%
  \BibitemOpen
  \bibfield  {author} {\bibinfo {author} {\bibfnamefont {F.}~\bibnamefont
  {Mazzola}}, \bibinfo {author} {\bibfnamefont {J.~W.}\ \bibnamefont {Wells}},
  \bibinfo {author} {\bibfnamefont {R.}~\bibnamefont {Yakimova}}, \bibinfo
  {author} {\bibfnamefont {S.}~\bibnamefont {Ulstrup}}, \bibinfo {author}
  {\bibfnamefont {J.~A.}\ \bibnamefont {Miwa}}, \bibinfo {author}
  {\bibfnamefont {R.}~\bibnamefont {Balog}}, \bibinfo {author} {\bibfnamefont
  {M.}~\bibnamefont {Bianchi}}, \bibinfo {author} {\bibfnamefont
  {M.}~\bibnamefont {Leandersson}}, \bibinfo {author} {\bibfnamefont
  {J.}~\bibnamefont {Adell}}, \bibinfo {author} {\bibfnamefont
  {P.}~\bibnamefont {Hofmann}}, \ and\ \bibinfo {author} {\bibfnamefont
  {T.}~\bibnamefont {Balasubramanian}},\ }\href {\doibase
  10.1103/PhysRevLett.111.216806} {\bibfield  {journal} {\bibinfo  {journal}
  {Phys. Rev. Lett.}\ }\textbf {\bibinfo {volume} {111}},\ \bibinfo {pages}
  {216806} (\bibinfo {year} {2013})}\BibitemShut {NoStop}%
\bibitem [{\citenamefont {Valla}\ \emph {et~al.}(2007)\citenamefont {Valla},
  \citenamefont {Kidd}, \citenamefont {Yin}, \citenamefont {Gu}, \citenamefont
  {Johnson}, \citenamefont {Pan},\ and\ \citenamefont {Fedorov}}]{Valla2007}%
  \BibitemOpen
  \bibfield  {author} {\bibinfo {author} {\bibfnamefont {T.}~\bibnamefont
  {Valla}}, \bibinfo {author} {\bibfnamefont {T.~E.}\ \bibnamefont {Kidd}},
  \bibinfo {author} {\bibfnamefont {W.-G.}\ \bibnamefont {Yin}}, \bibinfo
  {author} {\bibfnamefont {G.~D.}\ \bibnamefont {Gu}}, \bibinfo {author}
  {\bibfnamefont {P.~D.}\ \bibnamefont {Johnson}}, \bibinfo {author}
  {\bibfnamefont {Z.-H.}\ \bibnamefont {Pan}}, \ and\ \bibinfo {author}
  {\bibfnamefont {A.~V.}\ \bibnamefont {Fedorov}},\ }\href {\doibase
  10.1103/PhysRevLett.98.167003} {\bibfield  {journal} {\bibinfo  {journal}
  {Phys. Rev. Lett.}\ }\textbf {\bibinfo {volume} {98}},\ \bibinfo {pages}
  {167003} (\bibinfo {year} {2007})}\BibitemShut {NoStop}%
\bibitem [{\citenamefont {Inosov}\ \emph {et~al.}(2007)\citenamefont {Inosov},
  \citenamefont {Fink}, \citenamefont {Kordyuk}, \citenamefont {Borisenko},
  \citenamefont {Zabolotnyy}, \citenamefont {Schuster}, \citenamefont
  {Knupfer}, \citenamefont {B\"uchner}, \citenamefont {Follath}, \citenamefont
  {D\"urr}, \citenamefont {Eberhardt}, \citenamefont {Hinkov}, \citenamefont
  {Keimer},\ and\ \citenamefont {Berger}}]{Inosov2007}%
  \BibitemOpen
  \bibfield  {author} {\bibinfo {author} {\bibfnamefont {D.~S.}\ \bibnamefont
  {Inosov}}, \bibinfo {author} {\bibfnamefont {J.}~\bibnamefont {Fink}},
  \bibinfo {author} {\bibfnamefont {A.~A.}\ \bibnamefont {Kordyuk}}, \bibinfo
  {author} {\bibfnamefont {S.~V.}\ \bibnamefont {Borisenko}}, \bibinfo {author}
  {\bibfnamefont {V.~B.}\ \bibnamefont {Zabolotnyy}}, \bibinfo {author}
  {\bibfnamefont {R.}~\bibnamefont {Schuster}}, \bibinfo {author}
  {\bibfnamefont {M.}~\bibnamefont {Knupfer}}, \bibinfo {author} {\bibfnamefont
  {B.}~\bibnamefont {B\"uchner}}, \bibinfo {author} {\bibfnamefont
  {R.}~\bibnamefont {Follath}}, \bibinfo {author} {\bibfnamefont {H.~A.}\
  \bibnamefont {D\"urr}}, \bibinfo {author} {\bibfnamefont {W.}~\bibnamefont
  {Eberhardt}}, \bibinfo {author} {\bibfnamefont {V.}~\bibnamefont {Hinkov}},
  \bibinfo {author} {\bibfnamefont {B.}~\bibnamefont {Keimer}}, \ and\ \bibinfo
  {author} {\bibfnamefont {H.}~\bibnamefont {Berger}},\ }\href {\doibase
  10.1103/PhysRevLett.99.237002} {\bibfield  {journal} {\bibinfo  {journal}
  {Phys. Rev. Lett.}\ }\textbf {\bibinfo {volume} {99}},\ \bibinfo {pages}
  {237002} (\bibinfo {year} {2007})}\BibitemShut {NoStop}%
\bibitem [{\citenamefont {Basak}\ \emph {et~al.}(2009)\citenamefont {Basak},
  \citenamefont {Das}, \citenamefont {Lin}, \citenamefont {Nieminen},
  \citenamefont {Lindroos}, \citenamefont {Markiewicz},\ and\ \citenamefont
  {Bansil}}]{Basak2009}%
  \BibitemOpen
  \bibfield  {author} {\bibinfo {author} {\bibfnamefont {S.}~\bibnamefont
  {Basak}}, \bibinfo {author} {\bibfnamefont {T.}~\bibnamefont {Das}}, \bibinfo
  {author} {\bibfnamefont {H.}~\bibnamefont {Lin}}, \bibinfo {author}
  {\bibfnamefont {J.}~\bibnamefont {Nieminen}}, \bibinfo {author}
  {\bibfnamefont {M.}~\bibnamefont {Lindroos}}, \bibinfo {author}
  {\bibfnamefont {R.~S.}\ \bibnamefont {Markiewicz}}, \ and\ \bibinfo {author}
  {\bibfnamefont {A.}~\bibnamefont {Bansil}},\ }\href {\doibase
  10.1103/PhysRevB.80.214520} {\bibfield  {journal} {\bibinfo  {journal} {Phys.
  Rev. B}\ }\textbf {\bibinfo {volume} {80}},\ \bibinfo {pages} {214520}
  (\bibinfo {year} {2009})}\BibitemShut {NoStop}%
\bibitem [{\citenamefont {Rienks}\ \emph {et~al.}(2014)\citenamefont {Rienks},
  \citenamefont {\"Arr\"al\"a}, \citenamefont {Lindroos}, \citenamefont {Roth},
  \citenamefont {Tabis}, \citenamefont {Yu}, \citenamefont {Greven},\ and\
  \citenamefont {Fink}}]{Rienks2014}%
  \BibitemOpen
  \bibfield  {author} {\bibinfo {author} {\bibfnamefont {E.~D.~L.}\
  \bibnamefont {Rienks}}, \bibinfo {author} {\bibfnamefont {M.}~\bibnamefont
  {\"Arr\"al\"a}}, \bibinfo {author} {\bibfnamefont {M.}~\bibnamefont
  {Lindroos}}, \bibinfo {author} {\bibfnamefont {F.}~\bibnamefont {Roth}},
  \bibinfo {author} {\bibfnamefont {W.}~\bibnamefont {Tabis}}, \bibinfo
  {author} {\bibfnamefont {G.}~\bibnamefont {Yu}}, \bibinfo {author}
  {\bibfnamefont {M.}~\bibnamefont {Greven}}, \ and\ \bibinfo {author}
  {\bibfnamefont {J.}~\bibnamefont {Fink}},\ }\href {\doibase
  10.1103/PhysRevLett.113.137001} {\bibfield  {journal} {\bibinfo  {journal}
  {Phys. Rev. Lett.}\ }\textbf {\bibinfo {volume} {113}},\ \bibinfo {pages}
  {137001} (\bibinfo {year} {2014})}\BibitemShut {NoStop}%
\bibitem [{\citenamefont {Graf}\ \emph {et~al.}(2007)\citenamefont {Graf},
  \citenamefont {Gweon}, \citenamefont {McElroy}, \citenamefont {Zhou},
  \citenamefont {Jozwiak}, \citenamefont {Rotenberg}, \citenamefont {Bill},
  \citenamefont {Sasagawa}, \citenamefont {Eisaki}, \citenamefont {Uchida},
  \citenamefont {Takagi}, \citenamefont {Lee},\ and\ \citenamefont
  {Lanzara}}]{Graf2007}%
  \BibitemOpen
  \bibfield  {author} {\bibinfo {author} {\bibfnamefont {J.}~\bibnamefont
  {Graf}}, \bibinfo {author} {\bibfnamefont {G.-H.}\ \bibnamefont {Gweon}},
  \bibinfo {author} {\bibfnamefont {K.}~\bibnamefont {McElroy}}, \bibinfo
  {author} {\bibfnamefont {S.~Y.}\ \bibnamefont {Zhou}}, \bibinfo {author}
  {\bibfnamefont {C.}~\bibnamefont {Jozwiak}}, \bibinfo {author} {\bibfnamefont
  {E.}~\bibnamefont {Rotenberg}}, \bibinfo {author} {\bibfnamefont
  {A.}~\bibnamefont {Bill}}, \bibinfo {author} {\bibfnamefont {T.}~\bibnamefont
  {Sasagawa}}, \bibinfo {author} {\bibfnamefont {H.}~\bibnamefont {Eisaki}},
  \bibinfo {author} {\bibfnamefont {S.}~\bibnamefont {Uchida}}, \bibinfo
  {author} {\bibfnamefont {H.}~\bibnamefont {Takagi}}, \bibinfo {author}
  {\bibfnamefont {D.-H.}\ \bibnamefont {Lee}}, \ and\ \bibinfo {author}
  {\bibfnamefont {A.}~\bibnamefont {Lanzara}},\ }\href {\doibase
  10.1103/PhysRevLett.98.067004} {\bibfield  {journal} {\bibinfo  {journal}
  {Phys. Rev. Lett.}\ }\textbf {\bibinfo {volume} {98}},\ \bibinfo {pages}
  {067004} (\bibinfo {year} {2007})}\BibitemShut {NoStop}%
\bibitem [{\citenamefont {Ou}\ \emph {et~al.}(2009)\citenamefont {Ou},
  \citenamefont {Zhao}, \citenamefont {Zhang}, \citenamefont {Xie},
  \citenamefont {Shen}, \citenamefont {Zhu}, \citenamefont {Yang},
  \citenamefont {Che}, \citenamefont {Luo}, \citenamefont {Chen}, \citenamefont
  {Arita}, \citenamefont {Shimada}, \citenamefont {Namatame}, \citenamefont
  {Taniguchi}, \citenamefont {Cheng}, \citenamefont {Tsuei},\ and\
  \citenamefont {Feng}}]{Ou2009}%
  \BibitemOpen
  \bibfield  {author} {\bibinfo {author} {\bibfnamefont {H.~W.}\ \bibnamefont
  {Ou}}, \bibinfo {author} {\bibfnamefont {J.~F.}\ \bibnamefont {Zhao}},
  \bibinfo {author} {\bibfnamefont {Y.}~\bibnamefont {Zhang}}, \bibinfo
  {author} {\bibfnamefont {B.~P.}\ \bibnamefont {Xie}}, \bibinfo {author}
  {\bibfnamefont {D.~W.}\ \bibnamefont {Shen}}, \bibinfo {author}
  {\bibfnamefont {Y.}~\bibnamefont {Zhu}}, \bibinfo {author} {\bibfnamefont
  {Z.~Q.}\ \bibnamefont {Yang}}, \bibinfo {author} {\bibfnamefont {J.~G.}\
  \bibnamefont {Che}}, \bibinfo {author} {\bibfnamefont {X.~G.}\ \bibnamefont
  {Luo}}, \bibinfo {author} {\bibfnamefont {X.~H.}\ \bibnamefont {Chen}},
  \bibinfo {author} {\bibfnamefont {M.}~\bibnamefont {Arita}}, \bibinfo
  {author} {\bibfnamefont {K.}~\bibnamefont {Shimada}}, \bibinfo {author}
  {\bibfnamefont {H.}~\bibnamefont {Namatame}}, \bibinfo {author}
  {\bibfnamefont {M.}~\bibnamefont {Taniguchi}}, \bibinfo {author}
  {\bibfnamefont {C.~M.}\ \bibnamefont {Cheng}}, \bibinfo {author}
  {\bibfnamefont {K.~D.}\ \bibnamefont {Tsuei}}, \ and\ \bibinfo {author}
  {\bibfnamefont {D.~L.}\ \bibnamefont {Feng}},\ }\href {\doibase
  10.1103/PhysRevLett.102.026806} {\bibfield  {journal} {\bibinfo  {journal}
  {Phys. Rev. Lett.}\ }\textbf {\bibinfo {volume} {102}},\ \bibinfo {pages}
  {026806} (\bibinfo {year} {2009})}\BibitemShut {NoStop}%
\bibitem [{\citenamefont {Zhang}\ \emph {et~al.}(2014)\citenamefont {Zhang},
  \citenamefont {Dong}, \citenamefont {Yao}, \citenamefont {Chen},
  \citenamefont {Zhang}, \citenamefont {Gu}, \citenamefont {Zhou},
  \citenamefont {Lu}, \citenamefont {Chen},\ and\ \citenamefont
  {Shi}}]{Zhang2014}%
  \BibitemOpen
  \bibfield  {author} {\bibinfo {author} {\bibfnamefont {B.-B.}\ \bibnamefont
  {Zhang}}, \bibinfo {author} {\bibfnamefont {S.-T.}\ \bibnamefont {Dong}},
  \bibinfo {author} {\bibfnamefont {S.-H.}\ \bibnamefont {Yao}}, \bibinfo
  {author} {\bibfnamefont {Y.~B.}\ \bibnamefont {Chen}}, \bibinfo {author}
  {\bibfnamefont {S.-T.}\ \bibnamefont {Zhang}}, \bibinfo {author}
  {\bibfnamefont {Z.-B.}\ \bibnamefont {Gu}}, \bibinfo {author} {\bibfnamefont
  {J.}~\bibnamefont {Zhou}}, \bibinfo {author} {\bibfnamefont {M.-H.}\
  \bibnamefont {Lu}}, \bibinfo {author} {\bibfnamefont {Y.-F.}\ \bibnamefont
  {Chen}}, \ and\ \bibinfo {author} {\bibfnamefont {Y.~G.}\ \bibnamefont
  {Shi}},\ }\href {\doibase 10.1063/1.4893324} {\bibfield  {journal} {\bibinfo
  {journal} {Appl. Phys. Lett.}\ }\textbf {\bibinfo {volume} {105}},\ \bibinfo
  {pages} {062408} (\bibinfo {year} {2014})}\BibitemShut {NoStop}%
\bibitem [{\citenamefont {Pandiyan}(2013)}]{Pandiyan2013}%
  \BibitemOpen
  \bibfield  {author} {\bibinfo {author} {\bibfnamefont {M.~S.}\ \bibnamefont
  {Pandiyan}},\ }\emph {\bibinfo {title} {Phase Diagram andControl of
  Thermoelectric Propertiesof Sodium Cobaltate}},\ \href
  {https://pure.royalholloway.ac.uk/portal/en/publications/phase-diagram-and-control-of-thermoelectric-properties-of-sodium-cobaltate(3248e8ee-ae8c-4f27-a078-574c4653b1c4).html}
  {Ph.D. thesis},\ \bibinfo  {school} {Department of Physics, University of
  London} (\bibinfo {year} {2013})\BibitemShut {NoStop}%
\bibitem [{\citenamefont {Singh}\ and\ \citenamefont
  {Kasinathan}(2007)}]{Singh2007}%
  \BibitemOpen
  \bibfield  {author} {\bibinfo {author} {\bibfnamefont {D.}~\bibnamefont
  {Singh}}\ and\ \bibinfo {author} {\bibfnamefont {D.}~\bibnamefont
  {Kasinathan}},\ }\href {\doibase 10.1007/s11664-007-0154-0} {\bibfield
  {journal} {\bibinfo  {journal} {J. Electron. Mater.}\ }\textbf {\bibinfo
  {volume} {36}},\ \bibinfo {pages} {736} (\bibinfo {year} {2007})}\BibitemShut
  {NoStop}%
\bibitem [{\citenamefont {Giannozzi}\ \emph {et~al.}(2009)\citenamefont
  {Giannozzi}, \citenamefont {Baroni}, \citenamefont {Bonini}, \citenamefont
  {Calandra}, \citenamefont {Car}, \citenamefont {Cavazzoni}, \citenamefont
  {Ceresoli}, \citenamefont {Chiarotti}, \citenamefont {Cococcioni},
  \citenamefont {Dabo}, \citenamefont {{Dal Corso}}, \citenamefont
  {de~Gironcoli}, \citenamefont {Fabris}, \citenamefont {Fratesi},
  \citenamefont {Gebauer}, \citenamefont {Gerstmann}, \citenamefont
  {Gougoussis}, \citenamefont {Kokalj}, \citenamefont {Lazzeri}, \citenamefont
  {Martin-Samos}, \citenamefont {Marzari}, \citenamefont {Mauri}, \citenamefont
  {Mazzarello}, \citenamefont {Paolini}, \citenamefont {Pasquarello},
  \citenamefont {Paulatto}, \citenamefont {Sbraccia}, \citenamefont {Scandolo},
  \citenamefont {Sclauzero}, \citenamefont {Seitsonen}, \citenamefont
  {Smogunov}, \citenamefont {Umari},\ and\ \citenamefont
  {Wentzcovitch}}]{QE-2009}%
  \BibitemOpen
  \bibfield  {author} {\bibinfo {author} {\bibfnamefont {P.}~\bibnamefont
  {Giannozzi}}, \bibinfo {author} {\bibfnamefont {S.}~\bibnamefont {Baroni}},
  \bibinfo {author} {\bibfnamefont {N.}~\bibnamefont {Bonini}}, \bibinfo
  {author} {\bibfnamefont {M.}~\bibnamefont {Calandra}}, \bibinfo {author}
  {\bibfnamefont {R.}~\bibnamefont {Car}}, \bibinfo {author} {\bibfnamefont
  {C.}~\bibnamefont {Cavazzoni}}, \bibinfo {author} {\bibfnamefont
  {D.}~\bibnamefont {Ceresoli}}, \bibinfo {author} {\bibfnamefont {G.~L.}\
  \bibnamefont {Chiarotti}}, \bibinfo {author} {\bibfnamefont {M.}~\bibnamefont
  {Cococcioni}}, \bibinfo {author} {\bibfnamefont {I.}~\bibnamefont {Dabo}},
  \bibinfo {author} {\bibfnamefont {A.}~\bibnamefont {{Dal Corso}}}, \bibinfo
  {author} {\bibfnamefont {S.}~\bibnamefont {de~Gironcoli}}, \bibinfo {author}
  {\bibfnamefont {S.}~\bibnamefont {Fabris}}, \bibinfo {author} {\bibfnamefont
  {G.}~\bibnamefont {Fratesi}}, \bibinfo {author} {\bibfnamefont
  {R.}~\bibnamefont {Gebauer}}, \bibinfo {author} {\bibfnamefont
  {U.}~\bibnamefont {Gerstmann}}, \bibinfo {author} {\bibfnamefont
  {C.}~\bibnamefont {Gougoussis}}, \bibinfo {author} {\bibfnamefont
  {A.}~\bibnamefont {Kokalj}}, \bibinfo {author} {\bibfnamefont
  {M.}~\bibnamefont {Lazzeri}}, \bibinfo {author} {\bibfnamefont
  {L.}~\bibnamefont {Martin-Samos}}, \bibinfo {author} {\bibfnamefont
  {N.}~\bibnamefont {Marzari}}, \bibinfo {author} {\bibfnamefont
  {F.}~\bibnamefont {Mauri}}, \bibinfo {author} {\bibfnamefont
  {R.}~\bibnamefont {Mazzarello}}, \bibinfo {author} {\bibfnamefont
  {S.}~\bibnamefont {Paolini}}, \bibinfo {author} {\bibfnamefont
  {A.}~\bibnamefont {Pasquarello}}, \bibinfo {author} {\bibfnamefont
  {L.}~\bibnamefont {Paulatto}}, \bibinfo {author} {\bibfnamefont
  {C.}~\bibnamefont {Sbraccia}}, \bibinfo {author} {\bibfnamefont
  {S.}~\bibnamefont {Scandolo}}, \bibinfo {author} {\bibfnamefont
  {G.}~\bibnamefont {Sclauzero}}, \bibinfo {author} {\bibfnamefont {A.~P.}\
  \bibnamefont {Seitsonen}}, \bibinfo {author} {\bibfnamefont {A.}~\bibnamefont
  {Smogunov}}, \bibinfo {author} {\bibfnamefont {P.}~\bibnamefont {Umari}}, \
  and\ \bibinfo {author} {\bibfnamefont {R.~M.}\ \bibnamefont {Wentzcovitch}},\
  }\href {http://www.quantum-espresso.org} {\bibfield  {journal} {\bibinfo
  {journal} {J. Phys.: Condens. Matter}\ }\textbf {\bibinfo {volume} {21}},\
  \bibinfo {pages} {395502 (19pp)} (\bibinfo {year} {2009})}\BibitemShut
  {NoStop}%
\bibitem [{\citenamefont {Perdew}\ \emph {et~al.}(1996)\citenamefont {Perdew},
  \citenamefont {Burke},\ and\ \citenamefont {Ernzerhof}}]{Perdew1996}%
  \BibitemOpen
  \bibfield  {author} {\bibinfo {author} {\bibfnamefont {J.~P.}\ \bibnamefont
  {Perdew}}, \bibinfo {author} {\bibfnamefont {K.}~\bibnamefont {Burke}}, \
  and\ \bibinfo {author} {\bibfnamefont {M.}~\bibnamefont {Ernzerhof}},\ }\href
  {\doibase 10.1103/PhysRevLett.77.3865} {\bibfield  {journal} {\bibinfo
  {journal} {Phys. Rev. Lett.}\ }\textbf {\bibinfo {volume} {77}},\ \bibinfo
  {pages} {3865} (\bibinfo {year} {1996})}\BibitemShut {NoStop}%
\end{thebibliography}%


%merlin.mbs apsrev4-1.bst 2010-07-25 4.21a (PWD, AO, DPC) hacked
%Control: key (0)
%Control: author (8) initials jnrlst
%Control: editor formatted (1) identically to author
%Control: production of article title (-1) disabled
%Control: page (0) single
%Control: year (1) truncated
%Control: production of eprint (0) enabled
\begin{thebibliography}{3}%
\makeatletter
\providecommand \@ifxundefined [1]{%
 \@ifx{#1\undefined}
}%
\providecommand \@ifnum [1]{%
 \ifnum #1\expandafter \@firstoftwo
 \else \expandafter \@secondoftwo
 \fi
}%
\providecommand \@ifx [1]{%
 \ifx #1\expandafter \@firstoftwo
 \else \expandafter \@secondoftwo
 \fi
}%
\providecommand \natexlab [1]{#1}%
\providecommand \enquote  [1]{``#1''}%
\providecommand \bibnamefont  [1]{#1}%
\providecommand \bibfnamefont [1]{#1}%
\providecommand \citenamefont [1]{#1}%
\providecommand \href@noop [0]{\@secondoftwo}%
\providecommand \href [0]{\begingroup \@sanitize@url \@href}%
\providecommand \@href[1]{\@@startlink{#1}\@@href}%
\providecommand \@@href[1]{\endgroup#1\@@endlink}%
\providecommand \@sanitize@url [0]{\catcode `\\12\catcode `\$12\catcode
  `\&12\catcode `\#12\catcode `\^12\catcode `\_12\catcode `\%12\relax}%
\providecommand \@@startlink[1]{}%
\providecommand \@@endlink[0]{}%
\providecommand \url  [0]{\begingroup\@sanitize@url \@url }%
\providecommand \@url [1]{\endgroup\@href {#1}{\urlprefix }}%
\providecommand \urlprefix  [0]{URL }%
\providecommand \Eprint [0]{\href }%
\providecommand \doibase [0]{http://dx.doi.org/}%
\providecommand \selectlanguage [0]{\@gobble}%
\providecommand \bibinfo  [0]{\@secondoftwo}%
\providecommand \bibfield  [0]{\@secondoftwo}%
\providecommand \translation [1]{[#1]}%
\providecommand \BibitemOpen [0]{}%
\providecommand \bibitemStop [0]{}%
\providecommand \bibitemNoStop [0]{.\EOS\space}%
\providecommand \EOS [0]{\spacefactor3000\relax}%
\providecommand \BibitemShut  [1]{\csname bibitem#1\endcsname}%
\let\auto@bib@innerbib\@empty
%</preamble>
\bibitem [{\citenamefont {Giannozzi}\ \emph {et~al.}(2009)\citenamefont
  {Giannozzi}, \citenamefont {Baroni}, \citenamefont {Bonini}, \citenamefont
  {Calandra}, \citenamefont {Car}, \citenamefont {Cavazzoni}, \citenamefont
  {Ceresoli}, \citenamefont {Chiarotti}, \citenamefont {Cococcioni},
  \citenamefont {Dabo}, \citenamefont {{Dal Corso}}, \citenamefont
  {de~Gironcoli}, \citenamefont {Fabris}, \citenamefont {Fratesi},
  \citenamefont {Gebauer}, \citenamefont {Gerstmann}, \citenamefont
  {Gougoussis}, \citenamefont {Kokalj}, \citenamefont {Lazzeri}, \citenamefont
  {Martin-Samos}, \citenamefont {Marzari}, \citenamefont {Mauri}, \citenamefont
  {Mazzarello}, \citenamefont {Paolini}, \citenamefont {Pasquarello},
  \citenamefont {Paulatto}, \citenamefont {Sbraccia}, \citenamefont {Scandolo},
  \citenamefont {Sclauzero}, \citenamefont {Seitsonen}, \citenamefont
  {Smogunov}, \citenamefont {Umari},\ and\ \citenamefont
  {Wentzcovitch}}]{QE-2009}%
  \BibitemOpen
  \bibfield  {author} {\bibinfo {author} {\bibfnamefont {P.}~\bibnamefont
  {Giannozzi}}, \bibinfo {author} {\bibfnamefont {S.}~\bibnamefont {Baroni}},
  \bibinfo {author} {\bibfnamefont {N.}~\bibnamefont {Bonini}}, \bibinfo
  {author} {\bibfnamefont {M.}~\bibnamefont {Calandra}}, \bibinfo {author}
  {\bibfnamefont {R.}~\bibnamefont {Car}}, \bibinfo {author} {\bibfnamefont
  {C.}~\bibnamefont {Cavazzoni}}, \bibinfo {author} {\bibfnamefont
  {D.}~\bibnamefont {Ceresoli}}, \bibinfo {author} {\bibfnamefont {G.~L.}\
  \bibnamefont {Chiarotti}}, \bibinfo {author} {\bibfnamefont {M.}~\bibnamefont
  {Cococcioni}}, \bibinfo {author} {\bibfnamefont {I.}~\bibnamefont {Dabo}},
  \bibinfo {author} {\bibfnamefont {A.}~\bibnamefont {{Dal Corso}}}, \bibinfo
  {author} {\bibfnamefont {S.}~\bibnamefont {de~Gironcoli}}, \bibinfo {author}
  {\bibfnamefont {S.}~\bibnamefont {Fabris}}, \bibinfo {author} {\bibfnamefont
  {G.}~\bibnamefont {Fratesi}}, \bibinfo {author} {\bibfnamefont
  {R.}~\bibnamefont {Gebauer}}, \bibinfo {author} {\bibfnamefont
  {U.}~\bibnamefont {Gerstmann}}, \bibinfo {author} {\bibfnamefont
  {C.}~\bibnamefont {Gougoussis}}, \bibinfo {author} {\bibfnamefont
  {A.}~\bibnamefont {Kokalj}}, \bibinfo {author} {\bibfnamefont
  {M.}~\bibnamefont {Lazzeri}}, \bibinfo {author} {\bibfnamefont
  {L.}~\bibnamefont {Martin-Samos}}, \bibinfo {author} {\bibfnamefont
  {N.}~\bibnamefont {Marzari}}, \bibinfo {author} {\bibfnamefont
  {F.}~\bibnamefont {Mauri}}, \bibinfo {author} {\bibfnamefont
  {R.}~\bibnamefont {Mazzarello}}, \bibinfo {author} {\bibfnamefont
  {S.}~\bibnamefont {Paolini}}, \bibinfo {author} {\bibfnamefont
  {A.}~\bibnamefont {Pasquarello}}, \bibinfo {author} {\bibfnamefont
  {L.}~\bibnamefont {Paulatto}}, \bibinfo {author} {\bibfnamefont
  {C.}~\bibnamefont {Sbraccia}}, \bibinfo {author} {\bibfnamefont
  {S.}~\bibnamefont {Scandolo}}, \bibinfo {author} {\bibfnamefont
  {G.}~\bibnamefont {Sclauzero}}, \bibinfo {author} {\bibfnamefont {A.~P.}\
  \bibnamefont {Seitsonen}}, \bibinfo {author} {\bibfnamefont {A.}~\bibnamefont
  {Smogunov}}, \bibinfo {author} {\bibfnamefont {P.}~\bibnamefont {Umari}}, \
  and\ \bibinfo {author} {\bibfnamefont {R.~M.}\ \bibnamefont {Wentzcovitch}},\
  }\href {http://www.quantum-espresso.org} {\bibfield  {journal} {\bibinfo
  {journal} {J. Phys.: Condens. Matter}\ }\textbf {\bibinfo {volume} {21}},\
  \bibinfo {pages} {395502 (19pp)} (\bibinfo {year} {2009})}\BibitemShut
  {NoStop}%
\bibitem [{\citenamefont {Perdew}\ \emph {et~al.}(1996)\citenamefont {Perdew},
  \citenamefont {Burke},\ and\ \citenamefont {Ernzerhof}}]{Perdew1996}%
  \BibitemOpen
  \bibfield  {author} {\bibinfo {author} {\bibfnamefont {J.~P.}\ \bibnamefont
  {Perdew}}, \bibinfo {author} {\bibfnamefont {K.}~\bibnamefont {Burke}}, \
  and\ \bibinfo {author} {\bibfnamefont {M.}~\bibnamefont {Ernzerhof}},\ }\href
  {\doibase 10.1103/PhysRevLett.77.3865} {\bibfield  {journal} {\bibinfo
  {journal} {Phys. Rev. Lett.}\ }\textbf {\bibinfo {volume} {77}},\ \bibinfo
  {pages} {3865} (\bibinfo {year} {1996})}\BibitemShut {NoStop}%
\bibitem [{\citenamefont {Zhang}\ \emph {et~al.}(2015)\citenamefont {Zhang},
  \citenamefont {Zhang}, \citenamefont {Dong}, \citenamefont {Lv},
  \citenamefont {Chen}, \citenamefont {Yao}, \citenamefont {Zhang},
  \citenamefont {Gu}, \citenamefont {Zhou}, \citenamefont {Guedes},
  \citenamefont {Yu},\ and\ \citenamefont {Chen}}]{Zhang2015}%
  \BibitemOpen
  \bibfield  {author} {\bibinfo {author} {\bibfnamefont {B.-B.}\ \bibnamefont
  {Zhang}}, \bibinfo {author} {\bibfnamefont {N.}~\bibnamefont {Zhang}},
  \bibinfo {author} {\bibfnamefont {S.-T.}\ \bibnamefont {Dong}}, \bibinfo
  {author} {\bibfnamefont {Y.}~\bibnamefont {Lv}}, \bibinfo {author}
  {\bibfnamefont {Y.~B.}\ \bibnamefont {Chen}}, \bibinfo {author}
  {\bibfnamefont {S.}~\bibnamefont {Yao}}, \bibinfo {author} {\bibfnamefont
  {S.-T.}\ \bibnamefont {Zhang}}, \bibinfo {author} {\bibfnamefont {Z.-B.}\
  \bibnamefont {Gu}}, \bibinfo {author} {\bibfnamefont {J.}~\bibnamefont
  {Zhou}}, \bibinfo {author} {\bibfnamefont {I.}~\bibnamefont {Guedes}},
  \bibinfo {author} {\bibfnamefont {D.}~\bibnamefont {Yu}}, \ and\ \bibinfo
  {author} {\bibfnamefont {Y.-F.}\ \bibnamefont {Chen}},\ }\href {\doibase
  10.1063/1.4928384} {\bibfield  {journal} {\bibinfo  {journal} {{AIP}
  Advances}\ }\textbf {\bibinfo {volume} {5}},\ \bibinfo {pages} {087111}
  (\bibinfo {year} {2015})}\BibitemShut {NoStop}%
\end{thebibliography}%
\end{document}

\title{Anomalous band renormalization due to high energy $kink$ in the colossal thermoelectric material K$_{0.65}$RhO$_2$}

\author{Susmita Changdar}
\affiliation{Condensed Matter Physics and Material Sciences Department, S. N. Bose National Centre for Basic Sciences, Kolkata, West Bengal-700106, India}
\author{G. Shipunov}
\affiliation{Leibniz Institute for Solid State Research, IFW Dresden, D-01171 Dresden, Germany}
\author{N. B. Joseph}
\affiliation{Solid State and Structural Chemistry Unit, Indian Institute of Science, Bangalore, Karnataka-560012, India}
\author{N. C. Plumb}
\affiliation{Swiss Light Source, Paul Scherrer Institute,  CH-5232 Villigen PSI, Switzerland.}
\author{M. Shi}
\affiliation{Swiss Light Source, Paul Scherrer Institute, CH-5232 Villigen PSI, Switzerland.}
\author{B. B\"uchner}
\affiliation{Leibniz Institute for Solid State Research, IFW Dresden, D-01171 Dresden, Germany}
\author{Awadhesh Narayan}
\affiliation{Solid State and Structural Chemistry Unit, Indian Institute of Science, Bangalore, Karnataka-560012, India}
\author{S.\ Aswartham}
\affiliation{Leibniz Institute for Solid State Research, IFW Dresden, D-01171 Dresden, Germany}
\author{S.\ Thirupathaiah}
\email{setti@bose.res.in}
\affiliation{Condensed Matter Physics and Material Sciences Department, S. N. Bose National Centre for Basic Sciences, Kolkata, West Bengal-700106, India}
\date{\today}

\maketitle

\subsection{Experimental Methods}
$\emph{Single crystal growth}$:- Single crystals of K$_x$RhO$_2$ were grown from the mixtures of K$_2$CO$_3$ and Rh$_2$O$_3$. The total charge mixture of 4.5 grams was placed in an alumina crucible and heated to the 1200$^{\circ}$C in a box furnace, after a dwelling time of 2 hours, the furnace is slowly cooled to 950$^{\circ}$C and later fast-cooled to the room temperature. Plate-like hexagonal-shaped single crystals were grown at the bottom of the crucible. Crystals were grown in layered morphology in hexagonal structure up to few mm$^2$ in size. Compositional analysis from EDX gives the phase with the stoichiometry K$_{0.65(2)}$RhO$_2$. As grown single crystals were crushed and measured with powder x-ray diffraction (XRD).

$\emph{ARPES measurements}$:- ARPES measurements were performed in Swiss Light Source (SLS) at the SIS beamline using a VG-Scienta R4000 electron analyzer. Photon energy was varied between 20 and 140 eV. Overall energy resolution was set between 15 and 25~meV depending on the photon energy. The angular resolution was fixed at $0.2^{\circ}$. Samples were cleaved $\textit{in situ}$ at a sample temperature of 15 K and the chamber vacuum was better than $5\times10^{-11}$ mbar during the measurements.

$\emph{DFT Calculations}$:- First principles calculations were carried out based on density functional theory as implemented in quantum espresso code~\cite{QE-2009}. We used the generalized gradient approximation (GGA) to the exchange-correlation functional as formulated by Perdew-Burke-Ernzerhof (PBE)~\cite{Perdew1996}. A plane wave cutoff of 820 eV was used along with Brillouin Zone sampling over a 8$\times$8$\times$2 k-mesh. Lattice parameters for the stoichiometric KRhO$_2$ were taken from Ref.~\cite{Zhang2015}. The non-stoichiometric system, K$_{0.5}$RhO$_2$, was then modelled by removing 50$\%$ K atoms from the unit cell. The structure was relaxed until the forces on each atom were less than 10$^{-5}$ eV/$\AA$ before performing the band structure calculations.

\begin{figure*}[htbp]
\centering
\includegraphics[width=\linewidth]{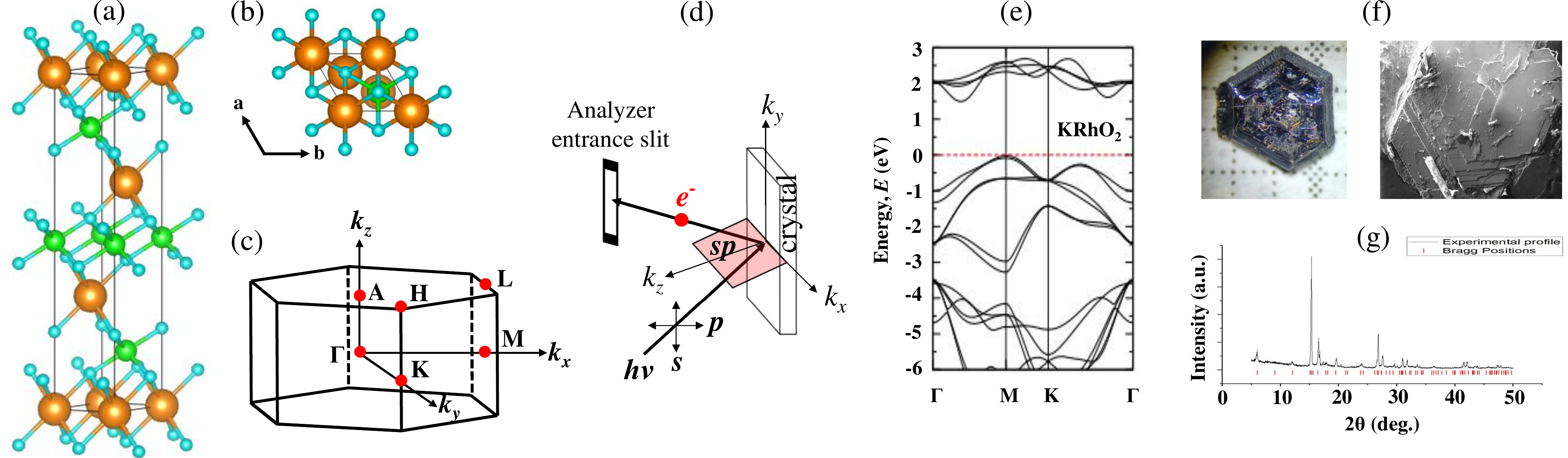}
\caption{(a) Crystal structure of KRhO$_2$. (b) Projected crystal structure onto the $ab$-plane. (c) Hexagonal Brillouin zone with identified high symmetry points. (d) ARPES measurement geometry. In (d), the $s$ and $p$ polarized lights are defined with respect to the scattering plane (sp) which is parallel to the $xz$ plane. (e) DFT band structure of the pristine KRhO$_2$. (f) As grown plate-like hexagonal crystals of K$_{0.65(2)}$RhO$_2$ (left) and SEM image (right). (g) XRD data of crushed K$_{0.65}$RhO$_2$ single crystals.}
\label{1}
\end{figure*}

\begin{figure*}[htbp]
\centering
\includegraphics[width=\linewidth]{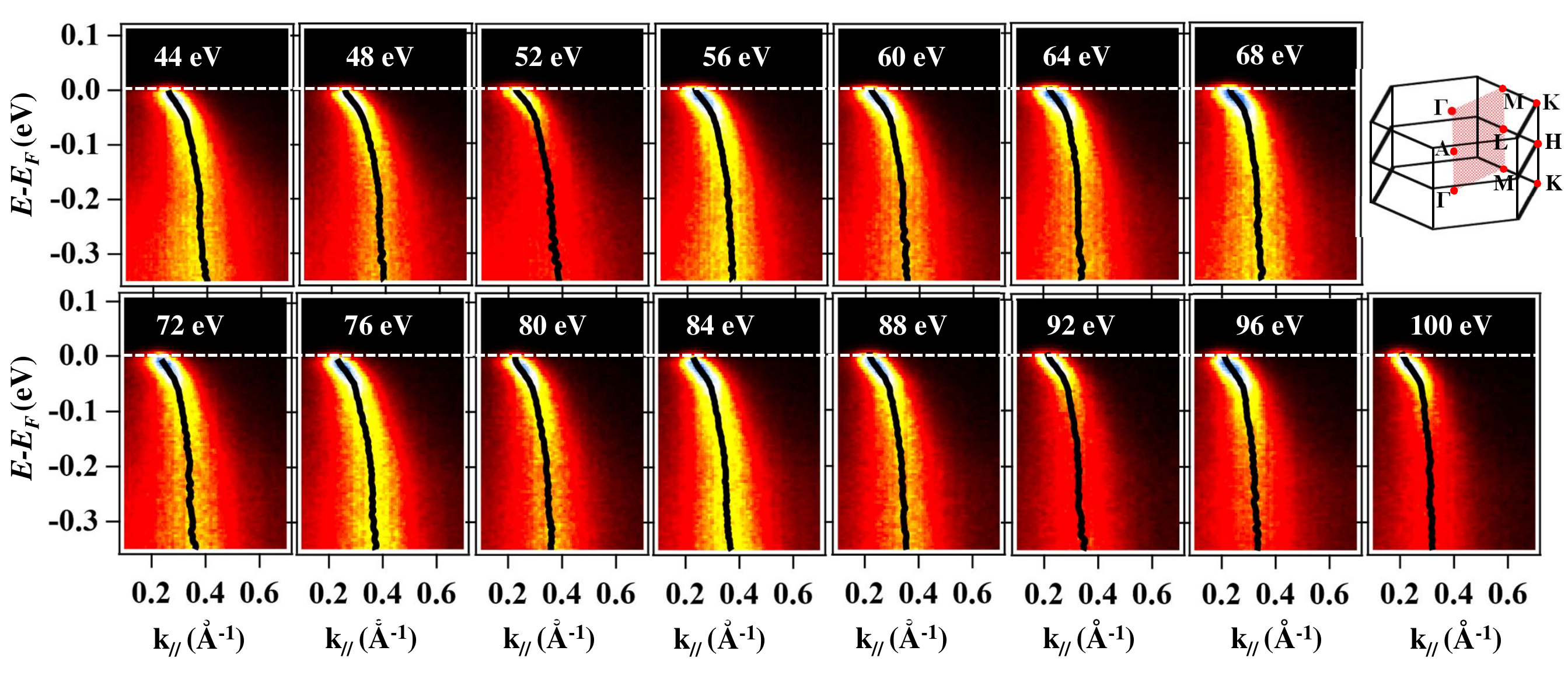}
\caption{Photon energy dependent ARPES data extracted in the $\Gamma M A L$ plane, as shown in the inset. Band dispersions extracted from the momentum distribution curves (MDCs) are overlapped onto the EDMs.}
\label{1}
\end{figure*}

\begin{figure*}[htbp]
\centering
\includegraphics[width=\linewidth]{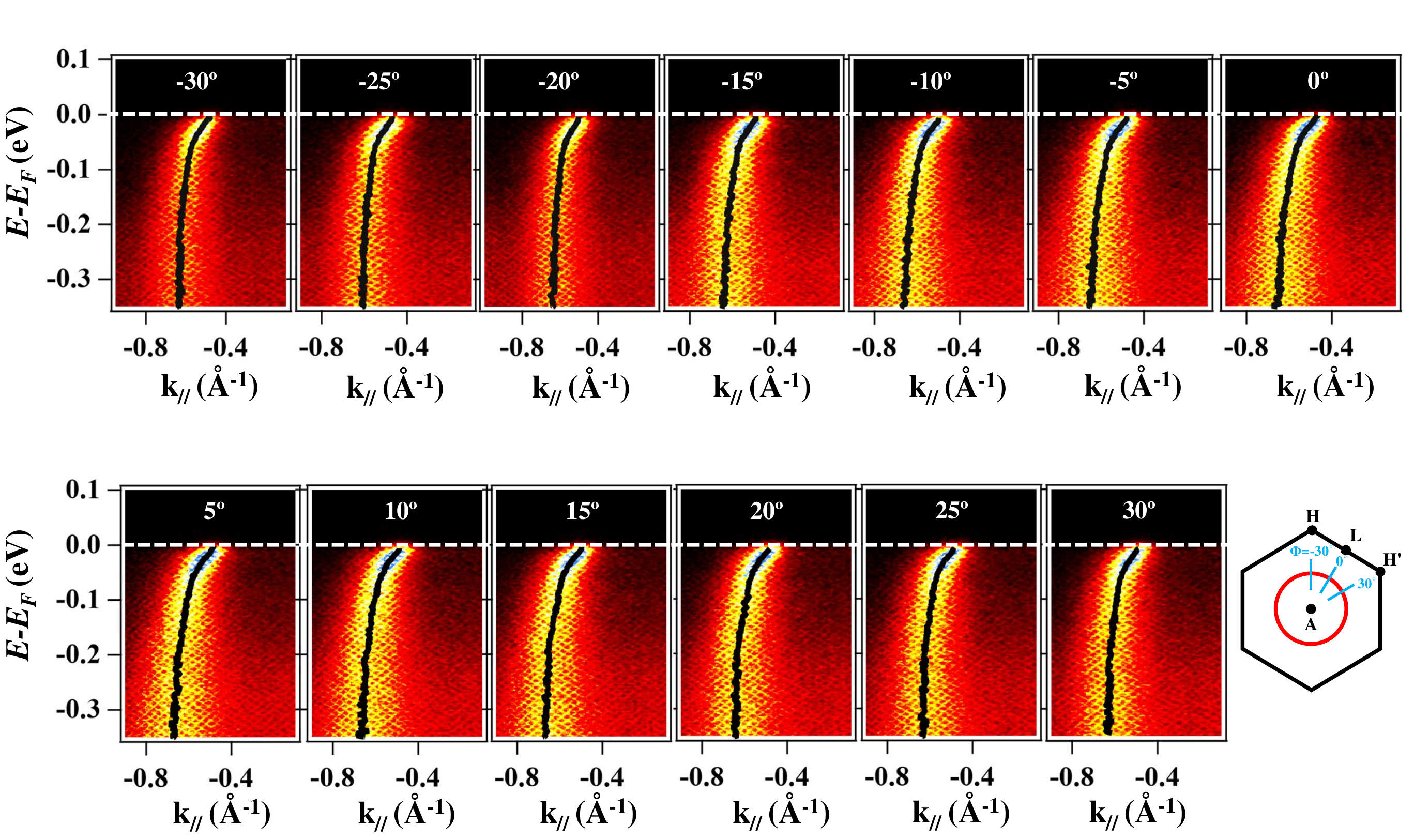}
\caption{In-plane momentum dependent ARPES data taken between $\Phi$=-30$^\circ$ and 30$^\circ$ as shown in the inset. Band dispersions extracted from the MDCs are overlapped onto the EDMs.}
\label{2}
\end{figure*}

\begin{figure*}[htbp]
\centering
\includegraphics[width=0.8\linewidth]{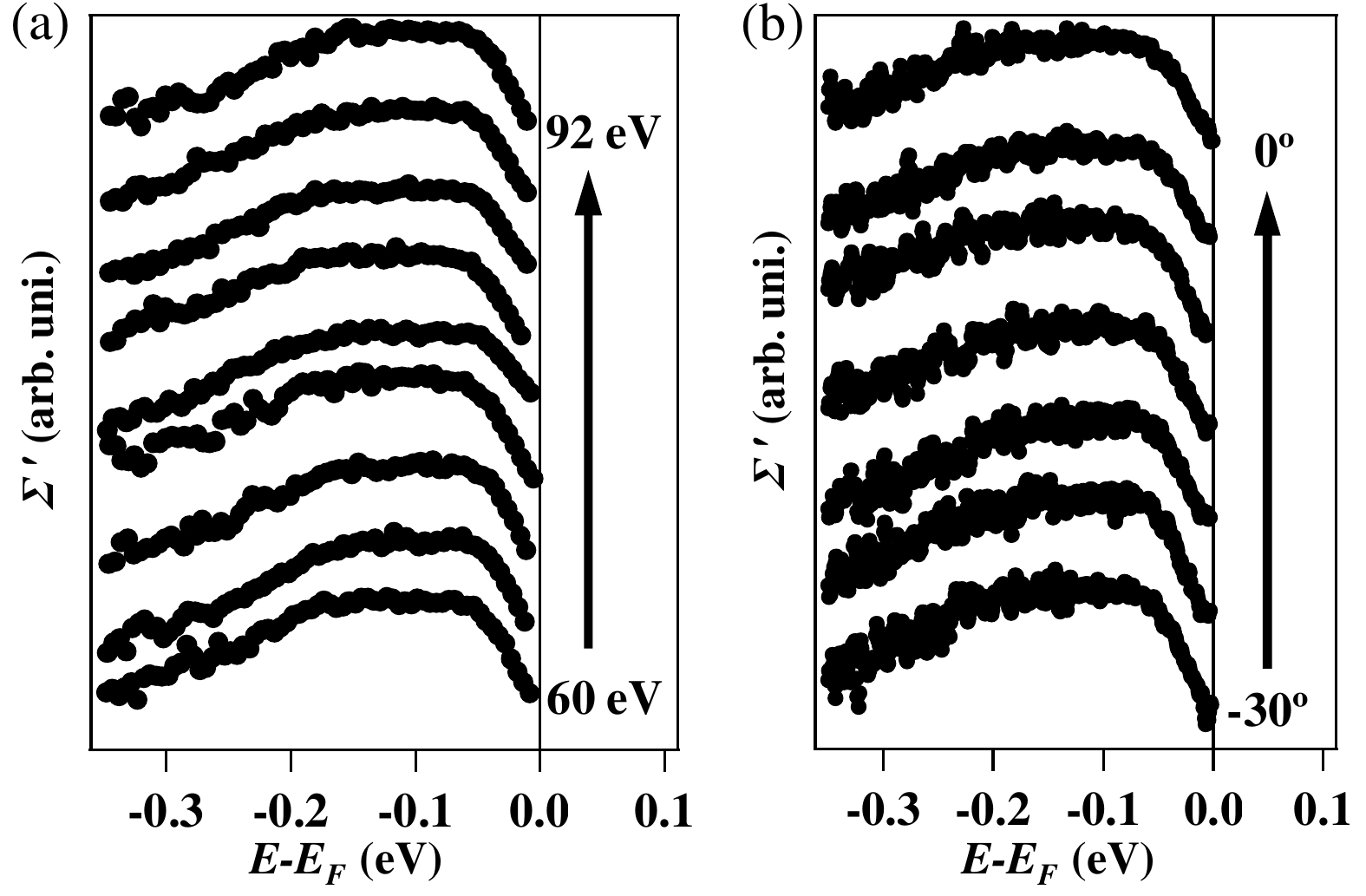}
\caption{Real part of self-energy ($\Sigma^{'} (E)$) taken as a function of photon energy (a) and $\Phi$ (b). For a better representation, the $\Sigma^{'} (E)$ data are stacked along the $y$-axis.}
\label{2}
\end{figure*}

\bibliography{supplement}